\documentclass[11pt]{article}
\usepackage{latexsym,cite,amssymb}
\makeatletter

\@addtoreset{equation}{section}
\makeatother

\newcommand\rpartial{{\stackrel{\leftarrow}{\partial}}}
\newcommand\lpartial{{\stackrel{\rightarrow}{\partial}}}

\newcommand{\diag}{{\rm diag}}

\newcommand{\hpartial}{{\hat{\partial}}}

\newcommand{\half}{{{\textstyle\frac{1}{2}}}}
\newcommand{\quarter}{{{\textstyle\frac{1}{4}}}}

\newcommand{\be}{\begin{equation} }
\newcommand{\ee}{\end{equation} }
\newcommand{\ba}{\begin{array}}
\newcommand{\ea}{\end{array}}

\newcommand{\su}{\mbox{su}}
\newcommand{\SU}{\mbox{SU}}
\newcommand{\so}{\mbox{so}}
\newcommand{\SO}{\mbox{SO}}

\def\cL{{\cal L}}
\def\cG{{\cal G}}
\def\cK{{\cal K}}
\def\cX{{\cal X}}

\def\tr{{\rm tr}}

\def\I_M{{I_{\scriptscriptstyle M\times M}}}

\def\YM{{\scriptscriptstyle \rm YM}}
\def\10D{{\scriptscriptstyle{\cN=16}}}
\def\6D{{\scriptscriptstyle{\cN=8}}}
\def\4D{{\scriptscriptstyle{\cN=4}}}
\def\3D{{\scriptscriptstyle{\cN=2}}}
\def\2D{{\scriptscriptstyle{\cN={1+1}}}}
\def\sDSYM{{\scriptscriptstyle{{6D}\,{\rm{SYM}}}}}
\def\fDSYM{{\scriptscriptstyle{{4D}\,{\rm{SYM}}}}}
\def\Mass{{\scriptscriptstyle \rm Massive}}
\def\BMN{{\scriptscriptstyle \rm BMN}}
\def\Myers{{\scriptscriptstyle \rm Myers}}
\def\cN{{\cal  N}}
\def\cM{{\cal  M}}
\def\cO{{\cal  O}}
\def\cP{{\cal P}}
\def\cV{{\cal V}}
\def\const{{\kappa}}
\def\cC{{\cal C}}
\def\mass{{\mu}}
\def\p{\cP}
\def\DYM{D_{{\scriptscriptstyle\rm YM}}}

\def\m{\hat{m}}
\def\T{{T}_{-}}

\def\dis{\displaystyle}
\newcommand\rmd{{\rm d}}

\newcommand\NambuGoto{{\rm \scriptscriptstyle{N.G.}}}
\newcommand\Polyakov{{\rm \scriptscriptstyle{Poly.}}}
\newcommand\particle{{\rm \scriptscriptstyle{particle}}}
\newcommand\PB{{\rm \scriptscriptstyle{P.B.}}}
\newcommand\NB{{\rm \scriptscriptstyle{N.B.}}}
\newcommand\MM{{\rm \scriptscriptstyle{M.M.}}}
\newcommand\typeI{{\rm \scriptscriptstyle{type\,I}}}
\newcommand\typeII{{\rm \scriptscriptstyle{type\,II}}}
\newcommand\pbrane{{{p{\scriptscriptstyle{-}}{\rm{brane}}}}}
\newcommand\mD{m}
\newcommand\A{A_{0}}
\newcommand\AG{A}
\newcommand\cc{{\rm{c.\,c.}}}

\begin{document}
\begin{titlepage}
\title{\vskip -60pt
{\small
\begin{flushright}
hep-th/0607005\\
MPP-2006-68
\end{flushright}}
\vskip 20pt Massive super Yang-Mills quantum mechanics\,:\\
classification and the relation to supermembrane\\
~}
\author{Nakwoo Kim${}^{\ast}$ and   Jeong-Hyuck Park${}^{\dagger}$}
\date{}
\maketitle
\vspace{-1.0cm}
\begin{center}
~~~\\
${}^{\ast}$Department of Physics and Research Institute of Basic
Science,
Kyung Hee University, Dongdaemun-gu, Seoul 130-701, Korea\\
~{}\\
${}^{\dagger}$Max-Planck-Institut f\"{u}r Physik,
F\"{o}hringer Ring 6,  80805  M\"{u}nchen,  Germany\\
~{}\\
{\small{Electronic correspondence: {{{nkim@khu.ac.kr}}},
{{{park@mppmu.mpg.de}}}}}\\
~~~\\
~~~\\
\end{center}
\begin{abstract}
\noindent We classify the supersymmetric mass  deformations of all
the super Yang-Mills quantum mechanics, which are obtained by
dimensional reductions of minimal super Yang-Mills in spacetime
dimensions: ten, six, four, three and two. The resulting actions can
be viewed as the matrix descriptions of  supermembranes in
nontrivial backgrounds  of one higher dimensional supergravity
theories. We also discuss the utmost generalization of the
light-cone formulation of the Nambu-Goto action for a $p$-brane,
including    time dependent  backgrounds.
\end{abstract}
\thispagestyle{empty}
\end{titlepage}
\newpage

\tableofcontents

\section{Introduction and summary}
Supersymmetry and gauge symmetry are  two principal symmetries   in
the Lagrangian description of a $p$-brane in superstring or
$\cM$-theory. Demanding both, the Lagrangian becomes considerably
constrained. For instance, when the Lagrangian contains only the
gauge multiplet, the action is uniquely given by the minimal super
Yang-Mills theory. As is well known, the minimal super Yang-Mills
theories exist not in arbitrary spacetime dimensions. They do  so
only in ten, six, four, three and two dimensions, which coincide
with the dimensions where the superstrings can be defined. The
constraints are due to  the requirement of the relevant Fierz
identity in order to implement the
non-Abelian gauge symmetry. \\

Dimensional reductions of the minimal super Yang-Mills field theories  lead
to one-dimensional gauge  theories, \textit{i.e.~}super Yang-Mills quantum
mechanics (SYMQM). In contrast to the rigidity of the formers,
the latters   allow \textit{mass
deformations}, without reducing the number of supersymmetries, as
discovered for the $\cN=16$~\cite{Berenstein:2002jq}, $\cN=8$~\cite{Kim:2002cr} and
$\cN={1+1}$~\cite{Park:2005pz} cases,
whose field theory origin can be traced in ten, six and two
dimensions,  respectively.
Throughout the paper, $\cN$ denotes the number of real dynamical  supersymmetries in
super Yang-Mills quantum mechanics.     Especially, the dimensional reduction of the
minimal super Yang-Mills in two-dimensions doubles the number of dynamical
supersymmetries~\cite{Park:2005pz},  whence our notation  `$\cN={1+1}$'. \\

In the present paper,
we classify, in a systematic way,   the mass deformations of all the
super Yang-Mills quantum mechanics with  $\cN=16,8,4,2,1+1$ supersymmetries.
Undeformed SYMQM are  given
 by the dimensional reductions of  the minimal super Yang-Mills field theories.
Utilizing the properties of the spinors and gamma matrices
in each case, we first consider adding generic mass terms for fermions and then
perform the supersymmetric completion.
Our final results are summarized in Table \ref{TheTable},
where `$\DYM$' denotes the spacetime dimension of the
corresponding minimal super Yang-Mills field theory, and $\mu$, $\mu_{1}$, $\mu_{2}$
are constant deformation  parameters,
while $\Lambda(t)$, $\rho(t)$ are arbitrary time dependent functions.\\

\begin{table}[htb]
\begin{center}
\begin{tabular}{lcccc}
\hline
{$\ba{c}\mbox{massive~SYM}\\ \mbox{quantum~\,mechanics}\ea$}
&$\DYM$&$\ba{c}\mbox{splitting~of}\\ \SO\!\left(\DYM-1\right)\ea$
& $\ba{c}\mbox{superalgebra}\ea$ &
${\ba{c}\mathrm{deformation}\\{\mathrm{parameter}}\ea}$\\
\hline
$~~~~~\cN=16$~~&~~ 10 ~~&~~$\SO(6)\times\SO(3)$~~\,&~~$\su(2|4)$& $\mu$\\
$~~~~~\cN=8$ type I ~~&~~ 6 ~~&~~$\SO(3)\times\SO(2)$~~~
&~~$\su(2|2)$ & $\mu$\\
$~~~~~\cN=8$ type II ~~&~~ 6 ~~&~~$\SO(4)$~~
&~~$\su(2|1)\oplus\su(2|1)$ & $\mu$\\
$~~~~~\cN=4$ type I ~~&~~  4 ~~&~~$\SO(3)~$~~&~~$\su(2|1)$ & $\mu_{1}$, $\mu_{2}$\\
$~~~~~\cN=4$ type II ~~&~~ 4 ~~&~~$\SO(2)$~~
&~~$\mbox{Clifford}_{4}({\mathbf{R}})$ & $\mu$\\
$~~~~~\cN=2$  ~~&~~ 3 ~~&~~$\SO(2)$~~
&~~$\mbox{Clifford}_{2}({\mathbf{R}})$& $\mu$\\
$~~~~~\cN={1+1}$ ~~&~~2~~&~~$\SO(1,2)$\,~~&~~
$\mbox{osp}(1|2,{\mathbf{R}})$~& $\Lambda(t)$, $\rho(t)$\\
\hline
\end{tabular}
~\\
\caption{Classification of massive super Yang-Mills quantum mechanics}
\end{center}
\label{TheTable}
\end{table}

For $\cN=16$ and $\cN=2$ cases
the deformations are unique, given by a single mass  parameter.
Both for $\cN=8$ and $\cN=4$ cases,
the deformations are two folds: type I and type II where only the former  contains a
Myers term~\cite{Myers:1999ps}.  In particular,    the number of
deformation parameters in  $\cN=4$ type I case is two, while in other cases it is one.
For  $\cN=1+1$ case, the deformation parameters are given by two arbitrary time
dependent functions, and accordingly  there are infinite number of deformation
parameters~\cite{Park:2005pz}.\\

Here we focus on the  deformations which
do not break any supersymmetry of the super Yang-Mills quantum
mechanics. We refer to \cite{Bonelli:2002mb} for the analysis on more generic
deformations of the $\cN=16$ SYMQM which break   supersymmetries partially. \\

Given the diversity of the  massive SYMQM we obtain,
it is natural to ask the physical origin of them.
So far, apart from our classification scheme,
three different ways of predicting or constructing massive super Yang-Mills
quantum mechanics are known:
(i) the light-cone description  of the  superparticle or supermembrane,
(ii) consistent truncations of the BMN matrix model caused by M5 branes, and (iii)
the compactification of four-dimensional super Yang-Mills on $S^{3}$.
In the following we
briefly discuss how our results may fit into  such physical insights.\\

As is well known,
the light-cone formulation~\cite{Susskind:1967rg,Kogut:1972di,deWit,Hoppe}
of the superparticle or supermembrane in the eleven-dimensional
flat background gives the undeformed $\cN=16$ SYMQM, which
is conjectured to describe the eleven-dimensional
$\cM$-theory in the flat background~\cite{Banks:1996vh,Susskind:1997cw,Sen,Seiberg}.
In 2002, Berenstein, Maldacena and Nastase (BMN) derived
a mass deformation of the $\cN=16$ SYMQM~\cite{Berenstein:2002jq}.
Their  derivation was based on the light-cone formulation of the
 superparticle  in  the maximally supersymmetric eleven-dimensional
pp-wave background~\cite{Figueroa-O'Farrill:2002ft,
Kowalski-Glikman,Figueroa-O'Farrill1,FO2}.
An alternative derivation is also available
from the supermembrane  aspect~\cite{Dasgupta:2002hx}.
The uniqueness of
the mass deformation of the $\cN=16$ SYMQM we show in the present paper
is consistent with the fact  that
the maximally supersymmetric backgrounds of the eleven-dimensional
supergravity are exhausted by the pp-wave  and
$AdS_{4}\times S^{7}$, $AdS_{7}\times S^{4}$ \cite{Figueroa-O'Farrill:2002ft}.\\

Clearly one can  expect  the same phenomena occur
 in lower dimensional supergravity theories: seven, five, four and three.
Namely, any supersymmetric pp-wave background therein  should give rise to
a novel massive SYMQM.  Indeed, the known maximally
supersymmetric plane wave solutions
in five- and four-dimensional  supergravity theories
(see \textit{e.g.}~\cite{Gauntlett:2002nw,Tod})
 correspond to our $\cN=4$ type II and $\cN=2$ massive SYMQM, respectively.
 In this spirit,  our $\cN=8$, $\cN=4$, $\cN=2$, $\cN={1+1}$
massive super Yang-Mills quantum mechanics
can be identified as $\cM$-theory matrix models in curved
backgrounds of the noncritical dimensions: seven, five, four and three,
respectively.\footnote{
For the discussion of noncritical $\cM$-theories see
\cite{Horava:2005tt,Horava:2005wm,Petkou:2005se,McGuigan3DM,Gomis:2005ce}. }  \\

Supersymmetric embedding of the  M2 and  M5 branes   into
the eleven-dimensional  pp-wave background has been analyzed  in \cite{Kim:2002tj}.
It turns out that there are two types of half BPS M5-branes, such that one
preserves $\SO(3)\times\SO(2)$ isometry of the transverse five-dimensional geometry
and the other preserves $\SO(4)$, whilst  the background pp-wave has the isometry
$\SO(3)\times\SO(6)$. This analysis consistently  matches with
our results: $\cN=8$ massive SYMQM type I and type II, respectively.
Namely, truncating any longitudinal degrees  of freedom (which should be described by
a self-dual two-form tensor), our two types of
$\cN=8$ massive super Yang-Mills quantum mechanics
describe the transverse degrees  of freedom of the M5-branes, treating
them  as `point-like' particles  (see
\cite{Kim:2002cr,Aharony:1997th,Aharony:1997an} for further discussion).\\

Another way of obtaining  massive SYMQM was proposed in \cite{Kim:2003rz},
to compactify the maximally supersymmetric
four-dimensional super Yang-Mills  on $S^{3}$ utilizing the
  superconformal invariance of the four-dimensional gauge theory~\cite{Nicolai:1988ek}.
The harmonic analysis of the dimensional reduction on $S^{3}$  produces a
quantum mechanical system  with an infinite tower of massive modes,
which can be arranged as irreducible representations of the
$\SO(4)\equiv\SU(2)\times \SU(2)$ isometry  of $S^{3}$.
When one keeps `half' of the lowest lying modes, the resulting quantum mechanical
system  is precisely the BMN matrix model~\cite{Kim:2003rz}.
The analysis was carefully revisited and generalized recently in
\cite{Ishiki:2006rt}.
It is obvious that one can apply this procedure to any other super Yang-Mills
theory in four-dimensions with less supersymmetries  and obtain a corresponding SYMQM.
From the super Yang-Mills with eight supercharges
one can reproduce   our  $\cN=8$  type I SYMQM, while from the
super Yang-Mills with four supercharges, \textit{i.e.}~pure super QCD,
one can  partially derive  our  $\cN=4$  type I SYMQM
with the restriction $\mass_{1}=0$.  It is worth while
to note that not all of the massive SYMQM in our classification
can be identified in the  manners discussed above.\\

The organization of the present paper is as follows: \\
In section \ref{REVISITED}, as a motivation to analyze the supersymmetric deformation of
each super Yang-Mills quantum mechanics, we review  the
light-cone formulation of the Nambu-Goto action for a generic $p$-brane.
In particular, we discuss the most general backgrounds which admit the light-cone
formulation.  The light-cone formulation converts, without any approximation,
the relativistic square root action
to a non-relativistic one where the kinetic term is simply velocity squared.
Especially for a membrane, the resulting action resembles the Yang-Mills action.
\\

In each of the following  sections
\ref{BMNSEC}, \ref{6DSEC}, \ref{4DSEC}, \ref{3DSEC},
\ref{2DSEC}, we discuss the  $\cN=16,\,8,\,4,\,2,\,{1+1}$
massive super Yang-Mills quantum mechanics separately.
We first briefly set up our conventions for the gamma matrices as well as the spinors,
such as $\su(2)$ Majorana-Weyl spinor in the $\cN=8$ case. We then
 present  explicitly the massive super Yang-Mills quantum mechanical systems
with the supersymmetry transformations of all the variables.
We further identify the corresponding super Lie algebras and write down the
maximally supersymmetric bosonic configurations. \\

Section \ref{COMMENTSEC} contains   some  comments.
The  appendix carries out    our  derivations of the most general
mass  deformations of all the super Yang-Mills  quantum mechanics. \\


\section{Light-cone formulation of a $p$-brane action  revisited\label{REVISITED}}
In this section, we review, with the utmost generalization,
the light-cone formulation of a $p$-brane action.
We first consider a relativistic point particle, and then generalize the analysis
to a  generic $p$-brane,  in view of applying to the  $p=2$ \textit{i.e.~}membrane case.
We take   Nambu-Goto action
 coupled to a ${p+1}$ form field as a relativistic description of a $p$-brane.
With an embedding of a $(p+1)$-dimensional worldvolume  into a
$D$-dimensional target spacetime
\be
x(\xi)~~:~~\xi^{\mu}~(0\leq\mu\leq p)
~\longrightarrow~x^{M}~(0\leq M\leq D-1)\,,
\ee
the action for a $p$-brane reads
\be
\ba{ll}
\displaystyle{S_{\pbrane}=\int\rmd^{p+1}\xi~\cL_{\pbrane}}\,,~~~~&~~~~
\cL_{\pbrane}=\cL_{\NambuGoto}+
\cL_{\scriptscriptstyle{C_{p+1}}}\,,
\ea
\ee
where the Lagrangian consists of two parts:
\be
\ba{ll}
\displaystyle{\cL_{\NambuGoto}=
-T\sqrt{-\det{\cG_{\mu\nu}}\,}\,,}~~~~&~~~~\displaystyle{
\cL_{\scriptscriptstyle{C_{p+1}}}=-\frac{1}{(p+1)!\,}\,
\epsilon^{\mu_{1}\mu_{2}\cdots\mu_{p+1}}\,\cC_{\mu_{1}\mu_{2}\cdots\mu_{p+1}}\,.}
\ea
\label{NG}
\ee
Here  $T$ is the $p$-brane tension of mass dimension $p+1$, while
$\cG_{\mu\nu}$ and  $\cC_{\mu_{1}\mu_{2}\cdots\mu_{p+1}}$ are respectively  the
induced metric and $p+1$ form gauge field on the worldvolume
\be
\ba{l}
\displaystyle{
\cG_{\mu\nu}(\xi)=\partial_{\mu}x^{M}\partial_{\nu}x^{N}G_{MN}(x)\,,}\\
{}\\
\displaystyle{
\cC_{\mu_{1}\mu_{2}\cdots\mu_{p+1}}(\xi)
=\partial_{\mu_{1}}x^{M_{1}}\partial_{\mu_{2}}x^{M_{2}}\cdots
\partial_{\mu_{p+1}}x^{M_{p+1}}C_{M_{1}M_{2}\cdots M_{p+1}}(x)\,.}
\ea
\ee
The Nambu-Goto Lagrangian $\cL_{\NambuGoto}$
measures the volume of the $p$-brane in the target spacetime,
and may be  replaced by a Polyakov action~:
\be
\displaystyle{\cL_{\Polyakov}=-\half T
\sqrt{-h\,}\Big[\,h^{-1\mu\nu}\partial_{\mu}x^{L}
\partial_{\nu}x^{M}G_{LM}(x)+1-p\,\Big]\,.}
\ee
Integrating out the auxiliary worldvolume metric $h_{\mu\nu}$,
using its equation of motion
\be
\ba{ll}
h_{\mu\nu}=\partial_{\mu}x^{L}\partial_{\nu}x^{M}G_{LM}(x)~~~~:~~~~\mbox{for~}p\neq 1\,,\\
{}&{}\\
h_{\mu\nu}\propto\partial_{\mu}x^{L}\partial_{\nu}x^{M}G_{LM}(x)
~~~~:~~~~\mbox{for~}p= 1\,,
\ea
\ee
one recovers $\cL_{\NambuGoto}$.
Henceforth we focus on the Nambu-Goto Lagrangian.\\

We show in the following subsections  that, for a certain class of backgrounds,
any sector of a  fixed  `light-cone momentum' can
be exactly described by a  `non-relativistic' action where  the kinetic term is simply
velocity squared.  An exact galilean invariance of
the light-cone formalism in a flat background
has been known for the point particle  by Susskind since 1967
\cite{Susskind:1967rg,Kogut:1972di}, and for the supermembrane
by de Wit, Hoppe and Nicolai in 1988 \cite{deWit} (for
generic $p$-branes see \cite{Hoppe}; also   for a non-light-cone formalism see
\cite{Erdmenger:2006eh}).   A new ingredient in the present review is an utmost
generalization to    nontrivial  time-dependent  backgrounds.\\

With  the light-cone coordinates
\be
x^{\pm}=\frac{1}{\sqrt{2}}\left(\pm x^{0}+x^{D-1}\right)\,,
\ee
we focus on  a  $D$-dimensional target spacetime background  given by a
metric $G_{LM}$ and a ${p+2}$ form field strength $F_{(p+2)}=\rmd C_{(p+1)}$.
We require an isometry along the light-cone direction $x^{-}$,
\be
\ba{ll}
\dis{\frac{\,\partial G_{LM}}{\partial x^{-}}=0\,,}~~~~~&~~~~~
\dis{\frac{\,\partial F_{(p+2)}}{\partial x^{-}}=0\,,}
\ea
\ee
and further set certain components of $G_{LM}$, $F_{(p+2)}$ to vanish,
\be
\ba{llll}
G_{--}=0\,,~~~~&~~~~G_{-a}=0\,,~~~~&~~~~F_{-M_{1}M_{2}\cdots M_{p+1}}=0\,,~~~~&~~~~
F_{a_{1}a_{2}\cdots a_{p+2}}=0\,.
\ea
\label{constraintGF}
\ee
Here we denote the target spacetime coordinates by
$x^{M}=\left(x^{+},x^{-},y^{a}\right)$ where
$a,b$ indices are for a $d$-dimensional Euclidean subspace, running from $1$ to $d=D-2$.
In summary, they are   of  the form~:
\begin{equation}
\ba{l}
\displaystyle{\rmd s^{2}
=\AG(y,x^{+})\!\left[2\rmd x^{+}\rmd x^{-}-2V(y,x^{+})\rmd x^{+}\rmd x^{+}
+2J_{a}(y,x^{+})\rmd x^{+}\rmd y^{a}+g_{ab}(y,x^{+})\rmd y^{a}\rmd y^{b}
\right]\,,}\\
{}\\
\displaystyle{F_{(p+2)}=
\frac{1}{(p+1)!}\,F_{+a_{1}a_{2}\cdots a_{p+1}}(y,x^{+})\,
\rmd x^{+}\wedge\rmd y^{a_{1}}\wedge\cdots\wedge\rmd y^{a_{p+1}}\,.}
\ea
\label{Mmetric0}
\end{equation}
Especially, the only non-vanishing
components of the field strength are $F_{+a_{1}a_{2}\cdots a_{p+1}}(y,x^{+})$.
 This implies, from $\rmd F=0$ and the Poincar\'{e} lemma,  that
 there exists a $p$-form field $\cV_{(p)}(y,x^{+})$  depending  on $y,x^{+}$ only such that
\be
F_{+a_{1}a_{2}\cdots a_{p+1}}(y,x^{+})=\partial_{a_{1}}\cV_{a_{2}\cdots a_{p+1}}+
(-1)^{p}\partial_{a_{2}}\cV_{a_{3}\cdots a_{p+1}a_{1}}\,+\,\cdots\,+\,(-1)^{p}
\partial_{a_{p+1}}\cV_{a_{1}\cdots a_{p}}\,.
\label{Poincare}
\ee
These properties are crucial for the light-cone formulation we analyze below.
All the known pp-wave type solutions assume this form (quite often $A=1$). It is worth to note that
unlike the metric $G_{LM}(y,x^{+})$ and the field strength $F(y,x^{+})$,
the gauge field $C_{(p+1)}(x)$  may depend on the light-cone coordinate $x^{-}$.\\

\subsection{Point particle aspect}
In this subsection, we focus on a point particle ($p=0$) which propagates
in the background (\ref{Mmetric0}).
After a gauge choice to identify the worldline with a target spacetime
light-cone coordinate

\be
\tau=\xi^{0}\equiv x^{+}\,,
\ee
the action (\ref{NG}) reads
\be
\ba{l}
\displaystyle{S_{\particle}=\int\rmd \tau\Big(
\cL_{\NambuGoto}-C_{+}(x)-\dot{x}^{-}C_{-}(x)-\dot{y}^{a}C_{a}(x)\Big)\,,}\\
{}\\
\displaystyle{\cL_{\NambuGoto}=
-\mD\sqrt{-\AG(y,\tau)\Big(2\dot{x}^{-}-2V(y,\tau)+2J_{a}(y,\tau)\dot{y}^{a}
+g_{ab}(y,\tau)\dot{y}^{a}\dot{y}^{b}\Big)
\,}\,.}
\ea
\label{MDBIactionG}
\ee
Here $\mD$ is the mass of the particle, and
dot  denotes the  derivative with respect to the
worldline  coordinate $\tau$, such that the canonical momenta are
\be
\ba{l}
\displaystyle{p_{-}=\frac{\mD A}{\,\sqrt{
-A\left(2\dot{x}^{-}-2V+2J_{a}\dot{y}^{a}+g_{ab}\dot{y}^{a}\dot{y}^{b}\right)\,}\,}
-C_{-}=-\frac{\mD^{2}A}{\,\cL_{\NambuGoto}\,}-C_{-}\,,}\\
{}\\
\displaystyle{p_{a}=\frac{\mD A\left( g_{ab}\dot{y}^{b}+J_{a}\right)}{\,\sqrt{
-A\left(2\dot{x}^{-}-2V+2J_{c}\dot{y}^{c}+g_{cd}\dot{y}^{c}\dot{y}^{d}\right)\,}\,}
-C_{a}=\left(p_{-}+C_{-}\right)\left(g_{ab}\dot{y}^{b}+J_{a}\right)-C_{a}\,.}
\ea
\label{Mmomenta}
\ee
Inverting these,  we express the velocities in terms of the phase space variables
\be
\ba{ll}
\displaystyle{\dot{y}^{a}=\frac{\, \bar{g}^{ab}\p_{b}\,}{
\p_{-}}-\bar{J}^{a}\,,}~~~~&~~~~
\displaystyle{\dot{x}^{-}
=V+\half J_{a}\bar{J}^{a}
-\frac{\,\bar{g}^{ab}\p_{a}\p_{b}
+m^{2}A\,}{2\p_{-}^{2}}\,,}
\ea
\label{Mvelocities}
\ee
where we define `modified momenta' $\p_{-}$, $\p_{a}$ by
\be
\ba{ll}
\p_{-}(p,x):=p_{-}+C_{-}(x^{-},y,\tau)\,,~~~~&~~~~
\p_{a}(p,x):=p_{a}+C_{a}(x^{-},y,\tau)\,,
\ea
\ee
and set $\bar{g}^{ab}$ to be the inverse of $g_{cd}$ such that
\be
\ba{lll}
\displaystyle{\bar{g}^{ab}g_{bc}=\delta^{a}{}_{c}\,,}~~~~&~~~~\dis{
\bar{J}^{a}(y,\tau):=\bar{g}^{ab}(y,\tau)J_{b}(y,\tau)\,,}~~~~&~~~~\dis{
J^{2}(y,\tau):=J_{a}(y\,\tau)\bar{J}^{a}(y,\tau)\,.}
\ea
\ee
Note that $\p_{-}$, $\p_{a}$ reduce to the canonical momenta $p_{-}$, $p_{a}$
in the absence of the one form field.   In terms of the former, the Hamiltonian reads
\be
\ba{ll}
\displaystyle{H}=&\dis{\frac{\,\bar{g}^{ab}(y,\tau)\p_{a}(p,x)
\p_{b}(p,x)
+m^{2}\AG(y,\tau)\,}{2\p_{-}(p,x)}
+C_{+}(x^{-},y,\tau)-\p_{a}(p,x)\bar{J}^{a}(y,\tau)}\\
{}&{}\\
{}&\dis{+\p_{-}(p,x)\Big(V(y,\tau)+\half{J}^{2}(y,\tau)\Big)\,.}
\ea
\label{MpHamiltonian}
\ee
The time evolution of the  modified  momenta $\p_{-}$, $\p_{a}$ are then,
from the Hamiltonian dynamics,
\be
\ba{l}
\displaystyle{\frac{\rmd \p_{-}}{\rmd \tau}=F_{+-}+F_{a-}
\frac{\partial H}{\,\partial p_{a}}\,,}\\
{}\\
\displaystyle{\frac{\rmd \p_{a}}{\rmd \tau}=F_{+a}+F_{-a}
\frac{\partial H}{\,\partial p_{-}}+F_{ba}
\frac{\partial H}{\,\partial p_{b}}-
\frac{\hpartial~}{\hpartial y^{a}}\big(H-C_{+}\big)
\,,}
\ea
\ee
where the  derivative with a hat $\hpartial$ denotes the partial derivative
with respect to  $(\p,x)$, rather than $(p,x)$.
In fact, all the partial derivatives above can be replaced by $\hpartial$, since in the
Hamiltonian (\ref{MpHamiltonian})
the canonical momenta appear only through the modified momenta.\\

From our general assumption (\ref{Mmetric0}),
we have $F_{-M}=F_{ab}=0$, and  there exists a function $\cV(y,\tau)$ satisfying
$\partial_{a}\cV=F_{+a}$ (\ref{Poincare}).  Thus, $\p_{-}$ is a constant of motion
\be
\displaystyle{\frac{\rmd \p_{-}}{\rmd \tau}=0\,,}
\ee
and for any sector with a fixed constant of motion $\p_{-}$,
the dynamics of the rest variables,
$y^{a}$, $1\leq a\leq d$, can be  described by the following Hamiltonian
\be
\ba{ll}
\dis{H^{-}\!\left(\p_{a},y^{b},\tau\right)
=}&\dis{\frac{\,\bar{g}^{ab}(y,\tau)\p_{a}\p_{b}+m^{2}\AG(y,\tau)\,}{
2\p_{-}}-\p_{a}\bar{J}^{a}(y,\tau)}\\
{}&{}\\
{}&\dis{\,+\p_{-}V(y,\tau)+\half\p_{-} J^{2}(y,\tau)-\cV(y,\tau)\,,}
\ea
\ee
such that
\be
\ba{ll}
\displaystyle{\frac{\rmd y^{a}}{\rmd \tau}=
\frac{~\hat{\partial}H^{{-}}}{\hat{\partial}\p_{a}}\,,}~~~~&~~~~
\displaystyle{\frac{\rmd \p_{a}}{\rmd \tau}=-
\frac{~\hat{\partial}H^{{-}}}{\hat{\partial}y^{a}}\,.}
\ea
\ee
Again,  $\hpartial$ denotes the partial derivative
with respect to  $(\p,x)$, rather than $(p,x)$.
Clearly, this Hamiltonian dynamics can be derived from the following `non-relativistic'
Lagrangian
\be
\displaystyle{\cL^{{-}}(y,\tau)=\p_{-}\Big[\half g_{ab}(y,\tau)
\dot{y}^{a}\dot{y}^{b}+{J}_{a}(y,\tau)\dot{y}^{a}
-V\left(y,\tau\right)-\half {\m^{2}}\AG(y,\tau)\Big]+\cV(y,\tau)\,,}
\label{MnonrelLG}
\ee
where  we set $\m:=m\p_{-}^{-1}$, and  $\p_{-}$ should be taken as a
$c$-number  rather than a dynamical variable.  From (\ref{Mmomenta}),
 $\p_{-}$ is always \textit{positive}. \\

$\cL^{-}$ is a `non-relativistic' Lagrangian in the sense that
the kinetic term in (\ref{MnonrelLG}) is
velocity squared. Nevertheless, the resulting dynamics matches perfectly with
the relativistic particle motion of a fixed constant of motion $\p_{-}$.\\

Now we  consider the dynamics of the  D0-brane gas.
We restrict  on the flat subspace, $g_{ab}=\delta_{ab}$. In this case,
the above non-relativistic Lagrangian for a single particle (\ref{MnonrelLG})
has a natural\footnote{
This is somewhat in contrast to the ordering ambiguity appearing in
non-Abelian generalizations of the relativistic action or  DBI action.
See \cite{non-Abelian} for some proposals to fix the ambiguity.  } generalization
to the Yang-Mills quantum mechanics, \textit{i.e.~}to the matrix model
with $\mbox{U}(N)$ gauge symmetry for the description of $N$  D-particles,
\be
\displaystyle{\cL^{-}_{\YM}=\p_{-}\tr\!\Big[\half D_{t}X^{a}D_{t}X_{a}+J_{a}
\left(X,\tau\right)D_{t}X^{a}\!
-V\left(X,\tau\right)-\half {\m^{2}}\! A\left(X,\tau\right)
+\,\cdots\,\Big]\!+\tr\!\Big[\cV\left(X,\tau\right)\!\Big].}
\label{MnonrelLGprime}
\ee
Here $X^{a}$ is a  Hermitian matrix of which the eigenvalues represent the positions
of the D-particles. Moreover, in (\ref{MnonrelLGprime}) the ordinary  time derivative
is replaced by the covariant time derivative involving
a non-dynamical gauge field $A_{0}$,
\be
\displaystyle{D_{t}X=\dot{X}-i[A_{0},X]\,.}
\label{covtd}
\ee
This allows  for the gauge symmetry,
\be
\ba{lll}
X~\longrightarrow~U^{-1}XU\,,~~~~&~~~~
A_{0}~\longrightarrow~U^{-1}A_{0}U+iU^{-1}\partial_{t}U\,,~~~~&~~~~
U\in\mbox{U}(N)\,.
\ea
\label{Ugauge}
\ee
The  equation of motion of the auxiliary gauge field  $A_{0}$  is
a secondary first-class constraint, and
the physical states are in the gauge singlet sector.\\

The reason  for the gauging is related to the identical property of the D-particles.
The diagonalization of the $X$ is not unique; the Weyl
group of $\mbox{U}(N)$ in (\ref{Ugauge}) acts by permuting the eigenvalues of $X$.
Physically, this reflects the fact that $D$-particles are identical
particles~\cite{Witten:1995im}. In  matrix models,
gauging the $\mbox{U}(N)$ symmetry  naturally
takes care of the ambiguity in the diagonalization.
Different diagonalizations correspond to the same
gauge orbit and hence to the same physical state~\cite{Park:2002eu}.
The gauging of the matrix model is also
consistent with the  gauge theory description of  $D$-brane dynamics.\\


The abbreviated part in (\ref{MnonrelLGprime}) corresponds to possible
terms which vanish when $N=1$ or the single particle case, such as
a commutator of the matrices.
Its explicit form can be uniquely determined by requiring both the maximal supersymmetry
and the gauge symmetry simultaneously,   as
we analyze in detail later.\\

\subsection{Membrane or $p$-brane aspect}
Here we generalize the above analysis on the point particle to  a generic $p$-brane
of which the dynamics is given by the Nambu-Goto action coupled to
a ${p+1}$ form (\ref{NG}).
Apparently the action is invariant under the $(p+1)$-dimensional worldvolume
diffeomorphism $\xi^{\mu}~\rightarrow~\xi^{\prime\mu}(\xi)$ provided
$x^{M}(\xi)$ transforms as a scalar
\be
x^{M}(\xi)~\longrightarrow~x^{\prime M}(\xi)=x^{M}(\xi^{\prime})\,.
\label{xMtrans}
\ee
This induces
\be
\dis{\cG_{\mu\nu}(\xi)~\longrightarrow~\cG_{\mu\nu}^{\prime}(\xi)=
\frac{\partial\xi^{\prime\kappa}}{\partial\xi^{\mu}}\,
\frac{\partial\xi^{\prime\lambda}}{\partial\xi^{\nu}}\,
\cG_{\kappa\lambda}(\xi^{\prime})\,.}
\ee

Below we will
break this gauge symmetry step by step \textit{via} some gauge fixing conditions
such that, at the end, only the
$p$-dimensional static volume preserving diffeomorphism will survive.\\

With the decomposition of the worldvolume coordinates into the temporal and spatial
parts
\be
\ba{ll}
\xi^{0}=\tau\,,~~~~&~~~~\xi^{i}=\sigma^{i}~~~(~1\leq i\leq p~)\,,
\ea
\ee
our first gauge fixing is to identify the worldvolume time
with a light-cone coordinate in the target spacetime,
\be
\tau\equiv x^{+}\,.
\ee
Now the unbroken  gauge symmetry is
\be
\ba{ll}
\tau~\longrightarrow~\tau^{\prime}=\tau\,,~~~~&~~~~
\sigma^{i}~\longrightarrow~\sigma^{\prime i}=f^{i}(\tau,\sigma)\,,
\ea
\ee
under which, in particular, $\cG_{\tau i}(\xi)$ component  transforms as
\be
\dis{\cG_{\tau i}(\xi)~\longrightarrow~\cG_{\tau i}^{\prime}(\xi)=
\frac{\partial f^{j}}{\partial\sigma^{i}}\,\cG_{jk}(\xi^{\prime})\left(
\partial_{\tau}f^{k}(\tau,\sigma)+\bar{\cG}^{kl}(\xi^{\prime})
\cG_{\tau l}(\xi^{\prime})\right)\,,}
\ee
where $\bar{\cG}^{kl}$ is the inverse of the ${p\times p}$ matrix $\cG_{ij}$
\textit{i.e.~}$\bar{\cG}^{ij}\cG_{jk}=\delta^{i}{}_{k}$.
With  arbitrary functions  $f^{k}(0,\sigma)$, $1\leq k\leq p$
at an initial time  $\tau=0$,
one can uniquely fix
its time evolution by demanding
\be
\partial_{\tau}f^{k}(\tau,\sigma)=-\bar{\cG}^{kl}\big(\tau,f(\tau,\sigma)\big)
\cG_{\tau l}\big(\tau,f(\tau,\sigma)\big)\,.
\ee
This  recurrently determines all the
higher order time derivatives of $f^{k}(\tau,\sigma)$.
Thus, it is possible to choose a gauge such that
\be
\forall~~~ 1\leq i\leq p\,,~~~~~\cG_{\tau i}=0\,.
\ee
At this stage
the unbroken gauge symmetry is the $p$-dimensional  worldvolume
`static'  diffeomorphism
\be
\ba{ll}
\tau~\longrightarrow~\tau^{\prime}=\tau\,,~~~~&~~~~
\sigma^{i}~\longrightarrow~\sigma^{\prime i}=f^{i}(\sigma)\,.
\ea
\label{static}
\ee
~\\

With the gauge choices above \textit{i.e.~}$\tau=x^{+}$ and $\cG_{\tau i}=0$,
the Nambu-Goto Lagrangian and the $p+1$ form Lagrangian  in
the target spacetime background  (\ref{Mmetric0}) read
\be
\ba{l}
\displaystyle{\cL_{\NambuGoto}=
-TA^{\frac{p+1}{2}}\sqrt{
\Big(-2\dot{x}^{-}+2V-2J_{a}\dot{y}^{a}
-g_{ab}\dot{y}^{a}\dot{y}^{b}\Big)
\det\!\Big({\partial_{i}y^{a}}{\partial_{j}y^{b}}g_{ab}\Big)\,}\,,}\\
{}\\
\displaystyle{\cL_{C_{p+1}}=
-\frac{1}{p!}\,\epsilon^{\,\tau j_{1}\cdots j_{p}}\,\partial_{j_{1}}x^{M_{1}}\cdots
\partial_{j_{p}}x^{M_{p}}\Big(C_{+M_{1}\cdots M_{p}}(x)
+\dot{x}^{-}C_{-M_{1}\cdots M_{p}}(x)
+\dot{y}^{a}C_{aM_{1}\cdots M_{p}}(x)\Big)\,,}
\ea
\label{NG2}
\ee
where the determinant is
for the $p\times p$ matrix, ${\partial_{i}y^{a}}{\partial_{j}y^{b}}g_{ab}
=\cG_{ij}$. The dynamical variables are
$x^{-}$ and $y^{a}$. \\

Especially for the light-cone variable $x^{-}$, we define a quantity $\p_{-}$ as
\be
\dis{\p_{-}:=\frac{\,\partial\cL_{\NambuGoto}}{\partial\dot{x}^{-}}=TA^{\frac{p+1}{2}}
\sqrt{\frac{
\det\!\Big({\partial_{i}y^{a}}{\partial_{j}y^{b}}g_{ab}(y,\tau)\Big)}{
~-2\dot{x}^{-}+2V-2J_{a}\dot{y}^{a}-g_{ab}\dot{y}^{a}\dot{y}^{b}~}}~~.}
\label{p-p}
\ee
We note that $\p_{-}$ is a scalar density with weight one such that
under the static diffeomorphism (\ref{xMtrans}), (\ref{static}), it transforms as
\be
\p_{-}(\tau,\sigma)
~\longrightarrow~\p_{-}^{\prime}(\tau,\sigma)=\left|{
\det\!\left(\frac{\partial\sigma^{\prime}}{\partial\sigma}\right)}\right|
\p_{-}(\tau,\sigma^{\prime})\,.
\ee
Hence, it is possible to choose a gauge such that at  $\tau=0$
the momentum has no spatial dependency,
\be
\dis{\forall~~
1\leq i\leq p\,,~~~~~\frac{\partial\p_{-}(0,\sigma)}{\partial\sigma^{i}}=0\,.}
\label{psigma}
\ee
This is the last of the gauge fixing conditions in our prescription.
The remaining  unbroken gauge symmetry is then  the
time-independent  $p$-dimensional volume-preserving diffeomorphism.  \\

Now we turn to the dynamics.
The Euler-Lagrangian equations of the $p$-brane action are
\be
\dis{\partial_{\mu}\!\left(\frac{\,\partial\cL_{\NambuGoto}}{
\partial\,\partial_{\mu}x^{m}}\right)-\frac{\,\partial\cL_{\NambuGoto}}{
\partial x^{m}}\,+\,\partial_{\mu}\!\left(\frac{\,\partial\cL_{C_{p+1}}}{
\partial\,\partial_{\mu}x^{m}}\right)-\frac{\,\partial\cL_{C_{p+1}}}{
\partial x^{m}}\,=\,0\,,}
\ee
where $x^{m}$ is either $x^{-}$ or $y^{a}$.
After straightforward manipulation one obtains
\be
\dis{\partial_{\mu}\!\left(\frac{\,\partial\cL_{C_{p+1}}}{
\partial\,\partial_{\mu}x^{m}}\right)-\frac{\,\partial\cL_{C_{p+1}}}{
\partial x^{m}}=\frac{1}{(p+1)!}\,\epsilon^{\mu_{1}\cdots\mu_{p+1}}\,
\partial_{\mu_{1}}x^{M_{1}}\cdots\partial_{\mu_{p+1}}x^{M_{p+1}}
\,F_{mM_{1}\cdots M_{p+1}}\,.}
\label{pformEL}
\ee
In particular, since $\cL_{\NambuGoto}$ depends on neither $x^{-}$ nor its spatial
derivatives $\partial_{j}x^{-}$,  the equation of motion of $x^{-}$ is
\be
\dis{\frac{\,\partial\p_{-}}{\partial{\tau}}=\frac{1}{\,p!}\,
\epsilon^{\,\tau j_{1}\cdots j_{p}}\,\partial_{j_{1}}x^{M_{1}}\cdots\partial_{j_{p}}
x^{M_{p}}\Big(F_{+-M_{1}\cdots M_{p}}+\dot{y}^{a}F_{a-M_{1}\cdots M_{p}}\Big)\,.}
\ee
From our main assumption (\ref{Mmetric0}), the right hand side vanishes and hence,
 $\p_{-}$ is time independent,
$\partial_{\tau}\p_{-}= 0$. Therefore along with the gauge choice (\ref{psigma}),
$~\p_{-}$ becomes   strictly a  constant on-shell,
\be
\dis{\forall~~0\leq \mu\leq p\,,~~~~~
\frac{\partial \p_{-}}{\partial\xi^{\mu}}=0\,.}
\ee
Regarding  the equations of motion of the remaining variables $y^{a}$, which read with
(\ref{Mmetric0})
\be
\dis{\frac{\partial~}{\partial{\tau}}\left(
\frac{\partial\cL_{\NambuGoto}}{\partial\,\dot{y}^{a}}\right)+\,
\frac{\partial~~}{\partial{\sigma^{i}}}\left(
\frac{\partial\cL_{\NambuGoto}}{\partial\,\partial_{i}{y}^{a}}\right)-
\frac{\partial\cL_{\NambuGoto}}{\partial{y}^{a}}
-\frac{1}{\,p!}\,
\epsilon^{\,\tau j_{1}\cdots j_{p}}\,\partial_{j_{1}}y^{b_{1}}\cdots\partial_{j_{p}}
y^{b_{p}}\,F_{+ab_{1}\cdots b_{p}}=0\,,}
\ee
we have\footnote{Surely, the derivative of the determinant in (\ref{yapre})
can be explicitly spelled out using the relation
$\delta\det M=M^{-1ab}\delta M_{ba}\det M$.
However, what is  more illuminating  expression in our analysis is the one
 in (\ref{yapre}).} for a fixed $c$-number $\p_{-}$,
\be
\ba{l}
\dis{\frac{\partial\cL_{\NambuGoto}}{\partial\,\dot{y}^{a}}=
\p_{-}\left(g_{ab}\dot{y}^{b}+J_{a}\right)\,,}\\
{}\\
\dis{\frac{\partial\cL_{\NambuGoto}}{\partial\,\partial_{i}{y}^{a}}=
-\left(\frac{\,T^{2}}{2\p_{-}}\right)A^{p+1}
\frac{\partial~~~}{\,\partial\,\partial_{i}{y}^{a}}
\det\!\Big({\partial_{j}y^{a}}{\partial_{k}y^{b}}g_{ab}(y,\tau)\Big)\,,}\\
{}\\
\dis{\frac{\partial\cL_{\NambuGoto}}{\partial{y}^{a}}=
\frac{\partial~~}{\,\partial{y}^{a}}\left[\p_{-}
\Big(\half g_{bc}\dot{y}^{b}\dot{y}^{c}+J_{b}\dot{y}^{b}-V\Big)
-\left(\frac{\,T^{2}}{2\p_{-}}\right)A^{p+1}
\det\!\Big({\partial_{j}y^{a}}{\partial_{k}y^{b}}g_{ab}\Big)\right]\,.}
\ea
\label{yapre}
\ee
Furthermore,  for the $p$-form $\cV$ satisfying (\ref{Poincare}), if we set
\be
\dis{\cL_{\cV}:=+\frac{1}{\,p!}\,\epsilon^{\tau j_{1}j_{2}\cdots j_{p}}
\partial_{j_{1}}y^{a_{1}}\partial_{j_{2}}y^{a_{2}}\cdots
\partial_{j_{p}}y^{a_{p}}\cV_{a_{1}a_{2}\cdots a_{p}}(y,\tau)\,,}
\ee
then in a similar fashion to (\ref{pformEL}) we obtain
\be
\dis{\partial_{i}\!\left(\frac{\,\partial\cL_{\cV}}{
\partial\,\partial_{i}y^{a}}\right)-\frac{\,\partial\cL_{\cV}}{
\partial y^{i}}=-\frac{1}{\,p!}\,\epsilon^{\tau j_{1}\cdots j_{p}}\,
\partial_{j_{1}}y^{b_{1}}\cdots\partial_{j_{p}}y^{b_{p}}
\,F_{+ab_{1}\cdots b_{p}}\,.}
\ee

Therefore, we conclude that \textit{for an arbitrary  sector  of the fixed
constant of motion $\p_{-}$,
the relativistic $p$-brane dynamics can be
exactly described by the following  `non-relativistic' action}\,:
\be
\ba{ll}
\cL^{{-}}=&\dis{\p_{-}\!\Big[\half g_{ab}\left(y,\tau\right)
\dot{y}^{a}\dot{y}^{b}+J_{a}(y,\tau)\dot{y}^{a}-V\left(y,\tau\right)
-\half\T^{2}\AG(y,\tau)^{p+1}
\det\!\Big({\partial_{i}y^{a}}{\partial_{j}y^{b}}g_{ab}(y,\tau)\!\Big)\Big]}\\
{}&{}\\
{}&\dis{\,+\frac{1}{\,p!}\,\epsilon^{j_{1}j_{2}\cdots j_{p}}\,
\partial_{j_{1}}y^{a_{1}}\cdots
\partial_{j_{p}}y^{a_{p}}\,\cV_{a_{1}\cdots a_{p}}(y,\tau)\,,}
\ea
\label{pnonrel}
\ee
where we set $\T:=T\p_{-}^{-1}$, and
from  (\ref{p-p}), \textit{$\p_{-}$ should be taken as a positive  $c$-number,
rather than a  dynamical  variable.}
The determinant in (\ref{pnonrel}) is for the $p\times p$ matrix.
 In the case $p=0$ \textit{i.e.~}a point particle, by setting
 the determinant to be $1$
 the Lagrangian consistently   reduces to the previous result (\ref{MnonrelLG}).
 For $p=1$ we obtain the usual  string action in the light-cone gauge.
After some scaling of the worldvolume coordinates we get
\be
\dis{\cL_{\rm{string}}=\half\!\left(
\partial_{\tau}{y}^{a}\partial_{\tau}{y}^{b}-\AG(y,\tau)^{2}
\partial_{\sigma}{y}^{a}\partial_{\sigma}{y}^{b}\right)g_{ab}\left(y,\tau\right)
+J_{a}(y,\tau)\dot{y}^{a}
-V\left(y,\tau\right)+{T^{-1}}\,
\partial_{\sigma}y^{a}\cV_{a}\,.}
\ee
For generic values of $p$, the determinant in
the above action (\ref{pnonrel}) can be expressed in terms of  the
Nambu bracket~\cite{Nambu:1973qe}
\be
\ba{l}
\dis{
\Big\{y^{a_{1}},y^{a_{2}},\cdots,y^{a_{p}}\Big\}_{\NB}:=
\epsilon^{j_{1}j_{2}\cdots j_{p}}\,\frac{\partial y^{a_{1}}}{\partial\sigma^{j_{1}}}
\frac{\partial y^{a_{2}}}{\partial\sigma^{j_{2}}}\cdots
\frac{\partial y^{a_{p}}}{\partial\sigma^{j_{p}}}\,,}\\
{}\\
\dis{\det\!\Big({\partial_{i}y^{a}}{\partial_{j}y^{b}}g_{ab}\Big)=
\frac{1}{p!}\Big\{y^{a_{1}},y^{a_{2}},\cdots,y^{a_{p}}\Big\}_{\NB}
\Big\{y^{b_{1}},y^{b_{2}},\cdots,y^{b_{p}}\Big\}_{\NB}
g_{a_{1}b_{1}}g_{a_{2}b_{2}}\cdots g_{a_{p}b_{p}}\,.}
\ea
\ee
The matrix regularization
is then to replace the dynamical fields
$y^{a}(\tau,\sigma)$ by
$N\times N$ Hermitian matrices $X^{a}(\tau)$ depending on the time only;
the ordinary time derivative by the covariant time derivative (\ref{covtd});
 and  further
the Nambu bracket by an anti-symmetrized matrix product,\footnote{
However, while the Nambu bracket satisfies the
generalized Jacobi identity
\[\dis{
\left\{\big\{f_{1},f_{2},\cdots,f_{p}\big\}_{\NB}\!,\,g_{2},\cdots,g_{p}\right\}_{\NB}
=\sum_{j=1}^{p}\,
\left\{f_{1},\cdots,f_{j-1},\big\{f_{j},g_{2},\cdots,g_{p}\big\}_{\NB}\!,\,f_{j+1},
\cdots,f_{p}\right\}_{\NB}\,,}
\]
the anti-symmetrized matrix product~(\ref{matrixreg}) does not do so except $p=2$ case.
See \cite{Curtright,Sheikh-Jabbari:2004ik} for related discussions.}
\be
\dis{\Big\{y^{a_{1}},y^{a_{2}},\cdots,y^{a_{p}}\Big\}_{\NB}~\Longleftrightarrow~
\left(\sqrt{-1}\right)^{\frac{1}{2}p(p-1)}
\Big[\,X^{a_{1}},X^{a_{2}},\cdots,X^{a_{p}}\,\Big]\,,}
\label{matrixreg}
\ee
where we set
\be
\dis{\Big[\,M_{1},M_{2},\cdots,M_{p}\,\Big]:=
\epsilon^{j_{1}j_{2}\cdots j_{p}}\,M_{j_{1}}M_{j_{2}}\cdots M_{j_{p}}\,.}
\ee
The  numerical factor in (\ref{matrixreg})
is chosen such that the right hand side is Hermitian provided
$X^{a}$'s are so. The resulting matrix model is then, for
$g_{ab}=\delta_{ab}$ (to avoid an ordering ambiguity),
\be
\ba{ll}
\multicolumn{2}{c}{\dis{S_{\MM}=\int{\rmd}\tau~\p_{-}\,\cL^{-}_{\MM}}\,,}\\
{}&{}\\
\dis{\cL^{-}_{\MM}=}&\tr\Big(
\dis{\half D_{t}X^{a}D_{t}X_{a}
+J_{a}\left(X,\tau\right)D_{t}X^{a}-V\left(X,\tau\right)}\Big)\\
{}&{}\\
{}&\dis{\,+\,\tr\!\left(
-\frac{\,\kappa_{p}^{2}\,}{\,2p!\,}\left({-1}\right)^{\frac{1}{2}p(p-1)}
\AG(X,\tau)^{p+1}\Big[\,X^{a_{1}},X^{a_{2}},\cdots,X^{a_{p}}\,\Big]^{2}\right)}\\
{}&{}\\
{}&\dis{\,+\,\tr\!\left(
\frac{\,\lambda_{p}}{\,p!\,}\left({\sqrt{-1}}\right)^{\frac{1}{2}p(p-1)}\,
\Big[\,X^{b_{1}},X^{b_{2}},
\cdots,X^{b_{p}}\Big]
\cV_{b_{1}b_{2}\cdots b_{p}}\left(X,\tau\right)\right)\,,}
\ea
\label{GeneralMM}
\ee
where $\kappa_{p},\lambda_{p}$ are constants. It is noteworthy  that
the first line  in the Lagrangian is universally present
irrespective of  $p$.\\

When $p=1$, \textit{i.e.~}for the string,
the matrix model essentially reduces to a $D-2$ copies of  harmonic oscillators
subject to  additional potentials $V$, $J_{a}$ and $\cV_{a}$; while
for $p=2$ \textit{i.e.~}membrane,  the Nambu bracket reduces to the Poisson bracket,
and the matrix regularization  can be justified in terms of the non-commutative
geometry as follows.
For a constant non-commutative deformation of a two-dimensional space
\be
\dis{\left[\sigma^{1},\sigma^{2}\right]=i\theta\,,}
\ee
the non-commutative geometry can be realized either by Moyal-Weyl
star product formalism on ordinary commutative space,
\be
\dis{f(\sigma)\star g(\sigma)=
f(\sigma)e^{i\frac{\theta}{2}\rpartial_{\!i}\epsilon^{ij}\lpartial_{\!j}}
g(\sigma)~~~\Longrightarrow~~~
\sigma^{1}\star\sigma^{2}-\sigma^{2}\star\sigma^{1}=i\theta\,,}
\ee
or equivalently by a matrix formalism generated by a pair of
$\infty\times\infty$ matrices satisfying the matrix commutator relation,
$\left[\hat{\sigma}^{1},\hat{\sigma}^{2}\right]=i\theta$.
The equivalence between the two formalisms follows from the isomorphism\footnote{
For a proof of the isomorphism
in the physics literature see \textit{e.g.~}\cite{Park:2003ku},
and for the
string theory aspect of the non-commutative geometry see \cite{Seiberg:1999vs}.
The isomorphism guarantees the associativity of the star product,
since the matrix product is associative.}
\begin{equation}
\cO(f)\cO(g)=\cO(f\star g)\,,\label{iso}
\end{equation}
where $\cO(f)$ is a Weyl ordering map from an ordinary commutative function
to a matrix, defined by
\begin{equation}
\dis{
\cO(\sigma^{j_{1}}\sigma^{j_{2}}\cdots \sigma^{j_{n}}):=
{\sum_{P=1}^{n!}\frac{1}{n!}}~
\hat{\sigma}^{{P_{1}}}\hat{\sigma}^{{P_{2}}}\cdots\hat{\sigma}^{{P_{n}}}\,.}
\end{equation}
Here $P$ denotes  the permutations of the $n$ indices $(j_{1},j_{2},\cdots,j_{n})$.

The justification of the matrix regularization  in the case of $p=2$ then follows
from an observation
that the Poisson bracket corresponds to the leading order term in the
start product commutator,
\be
\left[\,f,g\,\right]_{\star}=i\theta\left\{\,f,g\,\right\}_{\PB}\,+\,
O\!\left(\theta^{2}\right)\,.
\ee
~\\

The matrix model for a membrane contains terms of matrix commutator squared, and this
coincides with the potential in the Yang-Mills quantum mechanics.
In the rest of the paper, we analyze the most general
supersymmetric mass deformations of all the  super Yang-Mills
quantum mechanics,  without breaking any supersymmetry.
The resulting matrix models can be identified as the light-cone formulation of
the relativistic superparticle or supermembrane actions.\\
~\\
~\\


\section{$\cN=16$ super Yang-Mills quantum mechanics~: BMN matrix model\label{BMNSEC}}
In \cite{Figueroa-O'Farrill:2002ft}, Figueroa-O'Farrill and Papadopoulos classified
the maximally supersymmetric solutions of the eleven-dimensional supergravity theory.
Up to local isometry, $AdS_{4}\times S^{4}$, $AdS_{7}\times S^{7}$, pp-wave and
the flat spacetime exhaust them.  In particular, the pp-wave solution
reads~\cite{Figueroa-O'Farrill:2002ft,Kowalski-Glikman,Figueroa-O'Farrill1,FO2}
\begin{equation}
\ba{l}ds^{2}=2dx^{+}dx^{-}\!-\textstyle{\frac{1}{36}}\mu^{2}
\Big(x_{1}^{2}+\cdots+x_{6}^{2}+4x_{7}^{2}+4x_{8}^{2}+4x_{9}^{2}\Big)dx^{+}dx^{+}\!+
\displaystyle{\sum_{a=1}^{9}}\,dx^{a}dx^{a}\,,\\
{}\\
F_{789+}=\mu\,,
\ea
\label{ppw}
\end{equation}
where $\mu$ is a characteristic  mass parameter of the solution.
When $\mu=0$, the solution apparently  reduces to the flat background.\\

In \cite{Berenstein:2002jq}, Berenstein, Maldacena and Nastase (BMN) derived
the $\cM$-theory matrix model
in the above maximally  supersymmetric pp-wave background
\begin{equation}
 \ba{ll}
 \cL^{\10D}_{\BMN}=&\tr\Big(\half D_{t}X^{a}D_{t}X_{a}+\quarter [X^{a},X^{b}]^{2}
 +i\half\Psi^{\dagger}D_{t}\Psi-\half\Psi^{\dagger}\Gamma^{a}[X_{a},\Psi]\Big)\\
{}&{}\\
{}&+i\mu\,\tr\Big(
\textstyle{\frac{1}{8}}\Psi^{\dagger}\Gamma^{789}\Psi-X_{7}\left[X_{8},X_{9}\right]
\Big)\\
{}&{}\\
{}&-\textstyle{\frac{1}{72}}\mu^{2}\,\tr\Big(
X_{1}^{2}+X_{2}^{2}+X_{3}^{2}+X_{4}^{2}+X_{5}^{2}
+X_{6}^{2}+4X_{7}^{2}+4X_{8}^{2}+4X_{9}^{2}\Big)\,,
\ea
\label{BMN}
\end{equation}
where with $a=1,2,\cdots,9$,
$\Gamma^{a}=\left(\Gamma^{a}\right)^{\dagger}$ are Euclidean
nine-dimensional   gamma
matrices satisfying in general\footnote{For
simplicity one may   take the real gamma matrices such that $C=1$.}
\be
\ba{ll}
\left(\Gamma^{a}\right)^{T}=\left(\Gamma^{a}\right)^{\ast}=C^{-1}\Gamma^{a}C\,,
~~~~&~~~~C=C^{T}=\left(C^{\dagger}\right)^{-1}\,,
\ea
\ee
and $\Psi$ is a sixteen-component Majorana spinor $\Psi=C\Psi^{\ast}$.
The matrix model possesses 16 real dynamical supersymmetries (and hence $\cN=16$)~:
\be
\ba{l}
\delta A_{0}=i\Psi^{\dagger}\varepsilon(t)\,,~~~~~~~~~~~
\delta X^{a}=i\Psi^{\dagger}\Gamma^{a}\varepsilon(t)\,,\\
{}\\
\delta\Psi=\left(D_{t}X^{a}\Gamma_{a}-i\half [X^{a},X^{b}]\Gamma_{ab}-
\textstyle{\frac{1}{12}}\mu X\Gamma^{789}-
\textstyle{\frac{1}{4}}\mu \Gamma^{789}X\right)\varepsilon(t)\,,
\ea
\ee
where $X:=X^{a}\Gamma_{a}$, and the supersymmetric parameter is time dependent
\be
\varepsilon(t)=e^{\frac{1}{12}t\mass \Gamma^{789}}\varepsilon(0)\,.
\ee
As we show in Appendix \ref{derN=16},  BMN matrix model
is the \textit{unique} mass deformation of the
$\cN=16$ super Yang-Mills quantum mechanics.
The uniqueness  is consistent with the fact that,
among the maximally supersymmetric eleven-dimensional backgrounds,
only the pp-wave background (\ref{ppw}) can give the supersymmetric
mass deformation of the $\cN=16$ SYMQM  \textit{via} the
light-cone quantization.\\

The dynamical supersymmetries form a super Lie algebra $\su(2|4)$.
The classification of its supermultiplets was achieved
\cite{Kim:2002zg,Dasgupta:2002ru}, and  the corresponding classical
configurations were analyzed  in \cite{Park:2002cb} using the `projection' method
\cite{jhpBPS}.  For the
perturbative analysis on the spectrum, see  also \cite{Dasgupta:2002hx,Kim:2002if}.
In particular, the maximally supersymmetric configuration preserving all the
dynamical supersymmetries is given by a
static fuzzy sphere spanning the $7,8,9$ directions\,:
\be
\ba{lll}
{}\left[X_{p},X_{q}\right]=\textstyle{i\frac{1}{3}}\mass
\epsilon_{pqr}X^{r}\,,~~&~~D_{t}X^{p}=0\,,~~&~~
X^{1}=X^{2}=X^{3}=X^{4}=X^{5}=X^{6}=0\,,
\ea
\ee
where $p,q,r=7,8,9$ and $\epsilon_{789}=1$.
In addition, there are  16 real kinematical  supersymmetries
\begin{equation}
\ba{cc}
\delta A_{0}=\delta X^{a}=0\,, ~~~~&~~~~
\displaystyle{\delta\Psi=e^{-\frac{1}{4}t\mass \Gamma^{789}}\varepsilon^{\prime}\,.}
\ea
\end{equation}
~\\
\newpage


\section{$\cN=8$ super Yang-Mills quantum mechanics\label{6DSEC}}
In this section we analyze  the mass deformations of the
$\cN=8$ super Yang-Mills quantum mechanics which originates from the
six-dimensional minimal super Yang-Mills theory \textit{via} dimensional reduction.
We first review the six-dimensional  super Yang-Mills in order to set up our notations,
especially for the $\su(2)$ Majorana-Weyl spinor.
We then present the most general mass deformations of the
$\cN=8$ super Yang-Mills quantum mechanics.
It turns out that there exist two distinct types of  mass deformations:  type I and
type II,  with the corresponding superalgebra  $\su(2|2)$ and $\su(2|1)\oplus\su(2|1)$
respectively.
Only the former is
compatible  with the $\su(2)$ Majorana-Weyl condition,
while the latter  breaks the $\su(2)$ symmetry.
The latter is a rederivation of an  earlier work~\cite{Kim:2002cr}.
Here we simply present the results.
The detailed derivation is carried out in Appendix \ref{derivation6}.

\subsection{Minimal super Yang-Mills in six-dimensions\label{setup}}
In six-dimensional Minkowskian spacetime of the metric
$\eta=\diag(-+++++)$, the $8\times 8$
gamma matrices satisfy with $M=0,1,2,3,4,5$ \cite{Kugo:1982bn,Park:1998nr}
\be
\ba{lll}
\Gamma^{M}{}^{\dagger}=\Gamma_{M}=A\Gamma^{M}A^{\dagger}\,,
~~~~&~~~~A:=\Gamma^{12345}=A^{\dagger}=A^{-1}\,,\\
{}&{}\\
\Gamma^{M}{}^{T}=C\Gamma^{M}C^{\dagger}\,,
~~~~&~~~~C^{T}=-C\,,~~~~&~~~~C^{\dagger}=C^{-1}\,,\\
{}&{}\\
\Gamma^{M}{}^{\ast}=B\Gamma^{M}B^{\dagger}\,,
~~~~&~~~~B=CA=-B^{T}\,,~~~~&~~~~B^{\dagger}=B^{-1}\,.
\ea
\ee
The gamma ``seven" is given by
$\Gamma^{(7)}=\Gamma^{012345}$ which  satisfies
$\Gamma^{(7)}=\Gamma^{(7)\dagger}=\Gamma^{(7)-1}$ and
\be
\Gamma^{LMN}=\textstyle{\frac{1}{6}}\epsilon^{LMNPQR}\,\Gamma_{PQR}\Gamma^{(7)}\,,
\label{self-dual}
\ee
where $\epsilon^{012345}=+1$.\\

The $\mbox{su}(2)$ Majorana-Weyl spinor $\psi_{i}$, $i=1,2$, satisfies then
\be
\ba{ll}
\Gamma^{(7)}\psi_{i}=+\psi_{i}\,,~~~~~~~\bar{\psi}^{i}\Gamma^{(7)}=-\bar{\psi}^{i}
~~~~&~~~~:~~\mbox{chirality}\\
{}\\
\bar{\psi}^{i}=(\psi_{i})^{\dagger}A
=\epsilon^{ij}(\psi_{j})^{T}C~~~~&~~~~:~~\mbox{su(2)~Majorana}\,,
\ea
\label{su2MW}
\ee
where $\epsilon^{ij}$ is the usual  $2\times 2$ skew-symmetric unimodular matrix.  It is worth to note that
$\bar{\psi}^{i}\Gamma^{M_{1}M_{2}\cdots M_{{2n}}}\rho_{i}=0$ and
\be
\tr(i\bar{\psi}^{i}\Gamma^{M_{1}M_{2}\cdots M_{{2n+1}}}\rho_{i})=\left[\tr(i\bar{\psi}^{i}\Gamma^{M_{1}M_{2}\cdots M_{{2n+1}}}\rho_{i})\right]^{\dagger}=-(-1)^{n}\tr(i\bar{\rho}^{i}\Gamma^{M_{1}M_{2}\cdots M_{{2n+1}}}\psi_{i})\,,
\label{reality}
\ee
where $\psi_{i}$,  $\rho_{i}$ are two arbitrary   Lie algebra valued $\mbox{su}(2)$ Majorana-Weyl spinors.\\

The  six-dimensional super Yang-Mills Lagrangian  is
\begin{equation}
\dis{\cL_{\sDSYM}=\tr\left(-\textstyle{\frac{1}{4}}F_{LM}F^{LM}
-i\half\bar{\psi}^{i}\Gamma^{L}D_{L}\psi_{i}\right)\,,}
\label{6dsYML}
\end{equation}
where all the fields are in the adjoint representation of the gauge group such that,
with Hermitian Lie algebra valued gauge fields  $A_{M}$,
\be
\ba{ll}
D_{L}\psi_{i}=\partial_{L}\psi_{i}-i[A_{L},\psi_{i}]\,,~~~~&~~~~ F_{LM}=\partial_{L}A_{M}-\partial_{M}A_{L}-i[A_{L},A_{M}]\,.
\ea
\ee
From (\ref{reality}) the action  is real valued. \\

The supersymmetry transformations are,
with a $\mbox{su}(2)$ Majorana-Weyl supersymmetry parameter $\varepsilon_{i}$,
\be
\ba{ll}
\delta A_{M}=+i\bar{\varepsilon}^{i}\Gamma_{M}\psi_{i}=
-i\bar{\psi}^{i}\Gamma_{M}\varepsilon_{i}\,,~~~~&~~~~
\delta\psi_{i}=-\textstyle{\frac{1}{2}}F_{MN}\Gamma^{MN}\varepsilon_{i}\,.
\ea
\label{susy0}
\ee
In particular, $\delta\bar{\psi}^{i}=+\textstyle{\frac{1}{2}}
F_{MN}\bar{\varepsilon}^{i}\Gamma^{MN}$.
There are eight real supersymmetries.\\

The crucial Fierz identity for the supersymmetry invariance  is, with the chiral
projection matrix $P:=\half (1+\Gamma^{(7)})$,
\be
\left(\Gamma^{L}P\right)_{\alpha\beta}
\left(\Gamma_{L}P\right)_{\gamma\delta}+
\left(\Gamma^{L}P\right)_{\gamma\beta}
\left(\Gamma_{L}P\right)_{\alpha\delta}=0\,,
\label{Fierz}
\ee
which ensures the vanishing of the terms cubic in $\psi_{i}$
\be
\tr\!\left(\bar{\psi}^{i}\Gamma^{L}
[\delta A_{L},\,\psi_{i}]\right)=
\tr\!\left(\bar{\psi}^{i}\Gamma^{L}
[\,i\bar{\varepsilon}^{j}\Gamma_{L}\psi_{j}\,,\,\psi_{i}]\right)=0\,.
\label{Fierz2}
\ee
~\\

\subsection{Deformation  $\SO(5)\rightarrow\SO(3)\times\SO(2)$ : type I}
After a dimensional reduction of (\ref{6dsYML}) to the time $t$, we obtain
 a $\cN=8$ super Yang-Mills quantum mechanics,
containing  five Hermitian matrices   $X^{a}$, $a=1,2,3,4,5$,
\be
\cL^{\6D}_{0}=\tr\!\left(\half D_{t}X^{a}D_{t}X_{a}+\quarter [X^{a},X^{b}]^{2}
-i\half\bar{\psi}^{i}\Gamma^{t}D_{t}\psi_{i}
-\half\bar{\psi}^{i}\Gamma^{a}[X_{a},\psi_{i}]\right)\,.
\label{L06}
\ee
The mass dimensions are $1$ for the bosons and $\frac{3}{2}$
for the fermions so that the Lagrangian has mass dimension $4$.\\

We consider the most general
mass deformations of the above matrix model which preserves all
the supersymmetries.  From the chirality of the spinors,
the possible mass term for the fermion is of the form~:
\be
\ba{l}
\dis{\tr\!\left[\bar{\psi}\left(
M_{L}\Gamma^{L}-i\textstyle{\frac{1}{\,3!}}M_{abc}\Gamma^{abc}\right)
\psi\right]\,,}
\ea
\ee
where $M_{L}$, $M_{abc}$ are real parameters, which \textit{a priori}
may depend on time. However,
as we show in  Appendix (\ref{derivation6}), it turns out that in order to
admit $\cN=8$ supersymmetries they must be constants, and furthermore it is required
that  either $M_{L}=0$ or $M_{abc}=0$. Namely there exist two distinct types of
mass deformations:  type I and type II.  Only the former leads to
a mass deformation which is still compatible with the $\su(2)$ Majorana-Weyl condition.\\

With a constant mass parameter $\mu$,
$\cN=8$ type I massive super Yang-Mills quantum mechanics  reads
\be
\ba{ll}
\cL^{\6D}_{\typeI}=&\,\tr\Big(
\half D_{t}X^{a}D_{t}X_{a}+\quarter [X^{a},X^{b}]^{2}
-i\half\bar{\psi}^{i}\Gamma^{t}D_{t}\psi_{i}
-\half\bar{\psi}^{i}\Gamma^{a}[X_{a},\psi_{i}]\Big)\\
{}&{}\\
{}&+\,\tr\Big(i\textstyle{\frac{1}{8}}\mass\bar{\psi}^{i}\Gamma^{345}\psi_{i}
-i\mass[X_{3},X_{4}]X_{5}-\textstyle{\frac{1}{72}}\mass^{2}\left(
X_{1}^{2}+X_{2}^{2}+4X_{3}^{2}+4X_{4}^{2}+4X_{5}^{2}\right)\Big)\,.
\ea
\label{FINAL6}
\ee
The supersymmetry transformations are given by
\be
\ba{l}
\delta A_{0}=i\bar{\psi}^{i}\Gamma_{t}\varepsilon_{i}(t)\,,~~~~~~~
\delta X_{a}=i\bar{\psi}^{i}\Gamma_{a}\varepsilon_{i}(t)\,,\\
{}\\
\delta\psi_{i}=\Big(\Gamma^{ta}D_{t}X_{a}-i\half[X_{a},X_{b}]\Gamma^{ab}
-\textstyle{\frac{1}{12}}\mass X\Gamma^{345}-\textstyle{\frac{1}{4}}\mass
\Gamma^{345}X\Big)
\varepsilon_{i}(t)\,,
\ea
\label{SUSYf}
\ee
where
\be
\ba{ll}
X=\Gamma^{a}X_{a}\,,~~~~~&~~~~~
\varepsilon_{i}(t)=\dis{e^{\frac{1}{12}t\mass\,\Gamma^{t345}}\varepsilon_{i}(0)\,.}
\ea
\ee
Note that the Lagrangian manifestly possesses a
$\mbox{SO}(2)\times\mbox{SO}(3)\times\mbox{SU}(2)$
symmetry. The mass spectra  are
$\frac{1}{4}\mass$ for the fermions, $\frac{1}{6}\mass$ for the two  scalars and
$\frac{1}{3}\mass$ for the other three scalars.\\

\subsection{$\mbox{su}(2|2)$ superalgebra : type I}
Writing the Noether charge for the supersymmetry transformation (\ref{SUSYf}) as
\be
\tr\!\left(-i\bar{\psi}^{i}\Gamma^{t}\delta\psi_{i}\right):=i\bar{Q}^{i}\varepsilon_{i}(t)\,,
\ee
the supercharge satisfies the $\mbox{su}(2)$ Majorana-Weyl condition with the
opposite chirality  to the fermions $\varepsilon_{i}$, $\psi_{i}$,
\be
\ba{ll}
\Gamma^{(7)}Q_{i}=-Q_{i}\,,~~~~~~~\bar{Q}^{i}\Gamma^{(7)}=+\bar{Q}^{i}
~~~~&~~~~:~~\mbox{anti-chirality}\,,\\
{}\\
\bar{Q}^{i}=(Q_{i})^{\dagger}A
=\epsilon^{ij}(Q_{j})^{T}C~~~~&~~~~:~~\mbox{su(2)~Majorana}\,.
\ea
\label{su2MW2}
\ee
This is consistent with  the fact that the Noether charge is  real,
$i\bar{Q}^{i}\varepsilon_{i}(t)=-i\bar{\varepsilon}^{i}(t)Q_{i}
=\left(i\bar{Q}^{i}\varepsilon_{i}(t)\right)^{\dagger}$. \\

The supersymmetry algebra of the  $\cN=8$ type I massive SYMQM~(\ref{FINAL6})
reads explicitly
\be
\ba{ll}
~~~{}[H,Q_{i}]=+i\textstyle{\frac{1}{12}}\mass \Gamma_{12}Q_{i}\,,~~~~~~&~~~~~~
[H,\bar{Q}^{i}]=-i\textstyle{\frac{1}{12}}\mass \bar{Q}^{i}\Gamma_{12}\,,\\
{}&{}\\
\multicolumn{2}{l}{
\left\{Q_{i},\bar{Q}^{j}\right\}=2\delta_{i}^{~j}\Big(A(H-\textstyle{\frac{1}{6}}\mass
M_{12})-\textstyle{\frac{1}{6}}\mass\epsilon_{pqr} \Gamma^{p}M^{qr}\Big)P_{+}
+i\textstyle{\frac{2}{3}}\mass T_{i}{}^{j}\,\Gamma_{345}P_{+}\,,}
\ea
\label{susyalge}
\ee
and, as usual,
\be
\ba{ll}
{}[M_{12},Q_{i}]=+i\textstyle{\frac{1}{2}}\Gamma_{12} Q_{i}\,,
~~~~&~~~~[M_{12},\bar{Q}^{i}]=-i\textstyle{\frac{1}{2}}\bar{Q}^{i}\Gamma_{12}\,,\\
{}&{}\\
{}[M_{pq},Q_{i}]=+i\textstyle{\frac{1}{2}}\Gamma_{pq} Q_{i}\,,
~~~~&~~~~[M_{pq},\bar{Q}^{i}]=-i\textstyle{\frac{1}{2}}\bar{Q}^{i}\Gamma_{pq}\,,\\
{}&{}\\
\multicolumn{2}{l}{[M_{pq},M_{rs}]=i\left(\delta_{pr}M_{qs}-\delta_{ps}M_{qr}
-\delta_{qr}M_{ps}+\delta_{qs}M_{pr}\right)\,,}\\
{}&{}\\
{}[T_{i}{}^{j}, Q_{k}]=\delta^{j}{}_{k}Q_{i}-\half\delta_{i}{}^{j}Q_{k}\,,~~~~&~~~~
{}[T_{i}{}^{j},T_{k}{}^{l}]=\delta_{k}{}^{j}
T_{i}{}^{l}-\delta_{i}{}^{l}T_{k}{}^{j}\,,~~~~~~~T_{j}{}^{j}=0\,,\\
{}&{}\\
\multicolumn{2}{c}{[H,M_{12}]=0\,,~~~~~[H,M_{pq}]=0\,,~~~~~
[M_{12},M_{pq}]=0\,,~~~~~~[H,T_{i}{}^{j}]=0\,,~~~~~[M_{ab},T_{i}{}^{j}]=0\,.}\\
\ea
\ee
Here
$Q_{i}$, $H$, $M_{12}$, $M_{pq}$, $p,q=3,4,5$,
$T_{i}{}^{j}=\left(T_{j}{}^{i}\right)^{\dagger}$  refer to
the supercharges, Hamiltonian,  the  $\SO(3)\times\SO(2)$  generators,
the $\su(2)$ generator;
and  $P_{+}=\half\left(1+\Gamma^{(7)}\right)$ is the chiral projector.  \\

In particular,  we note that  ${{H-}\textstyle{\frac{1}{6}}\mass
M_{12}}$ is central\footnote{By a field redefinition~(\ref{rotating}),
one can  rewrite the Lagrangian such that the
new Hamiltonian itself is central.}
\be
{}[{{H-}\textstyle{\frac{1}{6}}\mass
M_{12}}\,,\,\mbox{anything}\,]=0\,,
\label{central}
\ee
and since $\left\{Q_{i},Q_{i}{}^{\dagger}\right\}$ is non-negative,
we have the unitary bound
\be
H\geq\textstyle{\frac{1}{6}}\mass M_{12}\,.
\label{unitaryb}
\ee
From the classification of the simple super Lie algebra by
Kac~\cite{Kac:1977em,Kac:1977qb},  we identify
the corresponding superalgebra\footnote{
Any inclusion of a brane charge in the simple super Lie algebra
(\textit{e.g.} \cite{Hyun:2002cm})  will inevitably lead  to  a
noncentral extension of the super Lie algebra~\cite{Lee:2004jx}.
This can be seen easily from
a Jacobi identity involving two supercharges $Q,\bar{Q}$ and a brane charge $Z$,
\[
\dis{\left[\left\{Q,\bar{Q}\right\},Z\right]=\left\{Q,\left[\bar{Q},Z\right]\right\}
+\left\{\bar{Q},\left[Q,Z\right]\right\}\,.}
\]
The left hand side corresponds to an infinitesimal rotation of the brane charge.
Since the brane charge is not singlet
under the rotations in general,
the left hand side does vanish. This  show that  the brane charge can not
commute with supercharges~\cite{Peeters:2003vz}.
In the present paper, we neglect the brane charges and
identify only the non-extended simple super Lie algebras
classified in \cite{Kac:1977em,Kac:1977qb}.} as a centrally extended
$\mbox{su}(2|2)$.\\

\subsection{$\cN=8$ supersymmetric configuration : type I}
From the supersymmetry transformation~(\ref{SUSYf}), it is straightforward to obtain the
following $8/8$ BPS equations \textit{i.e.~}the conditions for the
bosonic configuration to preserve all the eight supersymmetries,
\be
\ba{lll}
D_{t}X_{1}=-\textstyle{\frac{1}{6}}\mass X_{2}\,,~~~~&~~~
D_{t}X_{2}=+\textstyle{\frac{1}{6}}\mass X_{1}\,,~~~~&~~~D_{t}X_{p}=0\,,\\
{}&{}&{}\\
{}[X_{p},X_{q}]=i\textstyle{\frac{1}{3}}\mass\epsilon_{pqr}X^{r}\,,~~~~&~~~
[X_{i},X_{p}]=0\,,~~~~&~~~[X_{1},X_{2}]=0\,,
\ea
\ee
where $i=1,2$, $~p,q,r=3,4,5$,  and $\epsilon_{pqr}$ is  a
totally anti-symmetric tensor with $\epsilon_{345}=1$.
Note that the BPS equations imply  all the Euler-Lagrangian equations of the
$\cN=8$  super Yang-Mills quantum mechanics, including the Gauss
constraint.\footnote{However, in general, the BPS equations
for  less  supersymmetries do not necessarily imply
the Gauss constraint \cite{Park:2002cb}.}\\

The most general irreducible solution is given by
a fuzzy sphere spanning $(4,5,6)$ directions and rotating on the $(1,2)$ plane
\be
\ba{lll}
X_{1}=R\cos(\textstyle{\frac{1}{6}}t\mass)\,1\,,~~~&~~~
X_{2}=R\sin(\textstyle{\frac{1}{6}}t\mass)\,1\,,~~&~~~
X_{p}=\textstyle{\frac{1}{3}}\mass J_{p}\,,
\ea
\label{fuzzysphere}
\ee
where $R$ is the radius of the circular orbit on the $(1,2)$ plane, and
$J_{p}$, $p=4,5,6$ satisfy the standard  $\mbox{so}(3)$ commutator relations,
$[J_{p},J_{q}]=i\epsilon_{pqr}J^{r}$\,.
The rotating fuzzy sphere  saturates the unitary bound  (\ref{unitaryb}) as
$H=\textstyle{\frac{1}{6}}\mass M_{12}=\left(\frac{1}{6}\mass R\right)^{2}$.\\

\subsection{Deformation $\SO(5)\rightarrow\SO(4)$,
$~\su(2|1)\oplus\su(2|1)$ superalgebra : type II}
Type II mass deformation of the $\cN=8$ super Yang-Mills quantum mechanics breaks the
$\su(2)$ symmetry of the $\su(2)$ Majorana-Weyl spinors.
Hence, in this subsection we drop the $\su(2)$ index of the fermion
\be
\ba{ll}
\psi:=\psi_{1}\,,~~~~&~~~~\bar{\psi}=\psi^{\dagger}A=\bar{\psi}^{1}\,.
\ea
\ee
$\cN=8$ type II massive super Yang-Mills quantum mechanics then
reads~\cite{Kim:2002cr}
\be
\ba{ll}
\cL^{\6D}_{\typeII}=&\tr\Big(
\half D_{t}X^{a}D_{t}X_{a}+\quarter [X^{a},X^{b}]^{2}
-i\bar{\psi}\Gamma^{t}D_{t}\psi
-\bar{\psi}\Gamma^{a}[X_{a},\psi]\Big)\\
{}&{}\\
{}&+\tr\Big(\textstyle{\frac{1}{4}}\mu\bar{\psi}\Gamma^{1}\psi
-\textstyle{\frac{1}{72}}\mass^{2}\left(4X_{1}^{2}
+X_{2}^{2}+X_{3}^{2}
+X_{4}^{2}+X_{5}^{2}\right)\Big)\,.
\ea
\label{FINAL62}
\ee
The supersymmetry transformations are given by
\be
\ba{l}
\delta A_{0}=\bar{\psi}\Gamma_{0}\varepsilon(t)+
\bar{\varepsilon}(t)\Gamma_{0}\psi\,,
~~~~~~~\delta X_{a}=\bar{\psi}\Gamma_{a}\varepsilon(t)+
\bar{\varepsilon}(t)\Gamma_{a}\psi\,,\\
{}\\
\delta\psi=\Big(\!
-iD_{t}X_{a}\Gamma^{ta}-\half\left[X_{a},X_{b}\right]\Gamma^{ab}+
\textstyle{\frac{1}{4}}\mass\Gamma^{1}X+
\textstyle{\frac{1}{12}}\mass X\Gamma^{1}\Big)\varepsilon(t)\,,
\ea
\label{susymuII}
\ee
where
\be
\ba{ll}
X=\Gamma^{a}X_{a}\,,~~~~~&~~~~~
\varepsilon(t)=\dis{e^{-i\frac{1}{12}t\mass\,\Gamma^{t1}}\varepsilon(0)\,.}
\ea
\ee
Note that the Lagrangian manifestly possesses a $\mbox{SO}(4)$ symmetry.
The mass spectra  are  $\frac{1}{4}\mass$ for the fermions,
$\frac{1}{6}\mass$ for the four  scalars and $\frac{1}{3}\mass$ for the one scalar.\\

With  anti-chiral supercharges $Q=-\Gamma^{(7)}Q$,
$\bar{Q}=Q^{\dagger}A=\bar{Q}\Gamma^{(7)}$, the supersymmetry algebra corresponds to
$\su(2|1)\oplus\su(2|1)$,
\be
\ba{ll}
\multicolumn{2}{c}{
\left\{Q,\bar{Q}\right\}=2\Big(AH+i\textstyle{\frac{1}{12}}
\mass M_{mn}\Gamma^{mn1}-\textstyle{\frac{1}{12}}\mass
T\,\Gamma^{1}\Big)P_{+}\,,}\\
{}&{}\\
{}~~~~\left\{Q,Q\right\}=0\,,~~~~&~~~~[H,T\,]=0\,,\\
{}&{}\\
{}~~~~~[H,Q\,]=\textstyle{\frac{1}{12}}\mass {\Gamma^{t1}Q}\,,~~~~&~~~~
[H,\bar{Q}\,]=\textstyle{\frac{1}{12}}\mass \bar{Q}\Gamma^{t1}\,,\\
{}&{}\\
{}~~~~~[T,Q\,]=Q\,,~~~~&~~~~[T,\bar{Q}\,]=-\bar{Q}\,,
\ea
\label{susyalgep}
\ee
where  $M_{mn}$ corresponds to the $\so(4)=\su(2)\oplus\su(2)$ generators
for $(2,3,4,5)$ directions, and  $T$ is the $\mbox{u}(1)$ generator
of the phase rotation of the fermion. Unlike type I, in this case there is no
nontrivial $\cN=8$ supersymmetric configuration.\\
~\\
~\\

\section{$\cN=4$ super Yang-Mills quantum mechanics\label{4DSEC}}
In this section we analyze  the mass deformations of the
$\cN=4$ super Yang-Mills quantum mechanics which originates from the
four-dimensional minimal super Yang-Mills theory \textit{via} dimensional reduction.
After reviewing the four-dimensional  super Yang-Mills,
we  present the most general mass deformations of the
$\cN=4$ super Yang-Mills quantum mechanics.
Like $\cN=8$ case,  there exist two distinct types of  mass deformations:
type I and II.
Type I is a two-parameter family of deformations,
while type II contains only one parameter.

\subsection{Minimal  super Yang-Mills in four-dimensions\label{4DSYM}}
In four-dimensional Minkowskian spacetime of the metric $\eta=\diag(-+++)$, the $4\times 4$  gamma matrices satisfy with $\mu=0,1,2,3$
\be
\ba{lll}
\Gamma^{\mu}{}^{\dagger}=\Gamma_{\mu}=-A\Gamma^{\mu}A^{\dagger}\,,
~~~~&~~~~A=\Gamma^{t}=-A^{\dagger}\,,\\
{}&{}\\
\Gamma^{\mu}{}^{\ast}=+B\Gamma^{\mu}B^{\dagger}\,,
~~~~&~~~~B^{T}=B\,,~~~~&~~~~B^{\dagger}=B^{-1}\,,\\
{}&{}\\
\Gamma^{\mu}{}^{T}=-C\Gamma^{\mu}C^{\dagger}\,,
~~~~&~~~~C=-C^{T}=B\Gamma^{t}\,,~~~~&~~~~C^{\dagger}=C^{-1}\,.
\ea
\label{c41}
\ee
The spinors are then taken to meet  the  Majorana condition
\be
\ba{lll}
\bar{\psi}=\psi^{\dagger}\Gamma^{t}=\psi^{T}C
~~~&~~~\Longleftrightarrow~~~&~~~\psi^{\ast}=B\psi\,.
\ea
\label{c42}
\ee
The four-dimensional super Yang-Mills Lagrangian is
\begin{equation}
\dis{\cL_{\fDSYM}=\tr\left(-\textstyle{\frac{1}{4}}F_{\mu\nu}F^{\mu\nu}
-i\half\bar{\psi}\Gamma^{\mu}D_{\mu}\psi\right)\,.}
\label{c43}
\end{equation}
The supersymmetry transformations are
\be
\ba{ll}
\delta A_{\mu}=i\bar{\varepsilon}\Gamma_{\mu}\psi=-i\bar{\psi}\Gamma_{\mu}\varepsilon\,,~~~~&~~~~
\delta\psi=-\textstyle{\frac{1}{2}}F_{\mu\nu}\Gamma^{\mu\nu}\varepsilon\,.
\ea
\label{c44}
\ee
There are four real supersymmetries. \\

The  Fierz identity relevant to the supersymmetry invariance is
\be
(C\Gamma^{\mu})_{\alpha\beta}(C\Gamma_{\mu})_{\gamma\delta}+
(C\Gamma^{\mu})_{\beta\gamma}(C\Gamma_{\mu})_{\alpha\delta}+
(C\Gamma^{\mu})_{\gamma\alpha}(C\Gamma_{\mu})_{\beta\delta}=0\,.
\label{c45}
\ee
~\\

\subsection{Deformation  $\SO(3)\rightarrow\SO(3)$,
$~\su(2|1)$ superalgebra  : type I }
After a  dimensional reduction,  the four-dimensional  super Yang-Mills
gives a supersymmetric matrix model  containing
three Hermitian matrices,  $X^{a}$, $a=1,2,3$,
\be
\cL^{\4D}_{0}=\tr\!\left(\half D_{t}X^{a}D_{t}X_{a}+\quarter [X^{a},X^{b}]^{2}
-i\half\bar{\psi}\Gamma^{t}D_{t}\psi
-\half\bar{\psi}\Gamma^{a}[X_{a},\psi]\right)\,.
\label{L04}
\ee
As we show in Appendix \ref{derN=4}, there are two distinct ways of
deforming  this matrix model: type I and type II.
The former corresponds to a two-parameter $\mu_{1}$, $\mu_{2}$ family of deformation
\be
\ba{ll}
\cL^{\4D}_{\typeI}\!=&\tr\Big(\half D_{t}X^{a}D_{t}X_{a}+\quarter [X^{a},X^{b}]^{2}
-i\half\bar{\psi}\Gamma^{t}D_{t}\psi
-\half\bar{\psi}\Gamma^{a}[X_{a},\psi]\Big)\\
{}&{}\\
{}&+\,\tr\Big(
i\textstyle{\frac{1}{4}}
\mass_{1}\bar{\psi}\psi+i\textstyle{\frac{1}{4}}
\mass_{2}\bar{\psi}\Gamma^{123}\psi-i\mass_{2}\left[X_{1},X_{2}\right]X_{3}
-\textstyle{\frac{1}{18}}\left(\mu_{1}^{2}+\mu_{2}^{2}\right)X^{a}X_{a}\Big)\,.
\ea
\label{FINAL4}
\ee
Some characteristic features of the type I deformation are   the unbroken $\SO(3)$ symmetry and
the presence of the  Myers term.
The supersymmetry transformations are
\be
\ba{l}
\delta A_{0}=-i\bar{\psi}\Gamma_{t}\varepsilon(t)\,,~~~~~~~
\delta X_{a}=-i\bar{\psi}\Gamma_{a}\varepsilon(t)\,,\\
{}\\
\delta\psi=\Big(-\Gamma^{ta}D_{t}X_{a}+i\half[X_{a},X_{b}]\Gamma^{ab}
-\textstyle{\frac{1}{3}}\mass_{1}X
+\textstyle{\frac{1}{3}}\mass_{2} X\Gamma^{123} \Big)\varepsilon(t)\,,
\ea
\label{SUSYf4}
\ee
where
\be
\varepsilon(t)=\dis{e^{{\frac{1}{6}}
t\left(\mass_{1}\Gamma^{t}-\mass_{2}\Gamma^{t123}
\right)}
\varepsilon(0)\,.}
\ee
Diagonalizing the mass matrix
$\mass_{1}\Gamma^{t}+\mass_{2}\Gamma^{t123}$, we obtain the mass spectra,
$\textstyle{\frac{1}{2}}\sqrt{\mass_{1}^{2}+\mass_{2}^{2}}$ for the fermions and
$\frac{1}{3}\sqrt{\mass_{1}^{2}+\mass_{2}^{2}}$ for the bosons.\\

With the Majorana supercharge $\bar{Q}=Q^{T}C=Q^{\dagger}\Gamma^{t}$,
the corresponding supersymmetry algebra reads
\be
\ba{ll}
{}\left[H,Q\right]=-i\textstyle{\frac{1}{6}}
\left(\mu_{1}\Gamma^{t}+\mu_{2}\Gamma^{t123}\right)Q\,,~~&~~~
{\left\{Q,\bar{Q}\right\}=
2\Big(\Gamma^{t}H+\textstyle{\frac{1}{6}}\mu_{1}\Gamma^{ab}M_{ab}
-\textstyle{\frac{1}{6}}\mass_{2}\epsilon_{abc}\Gamma^{a}M^{bc}\Big).}
\ea
\label{fTISA}
\ee
This can be identified as   $\su(2|1)$ super Lie algebra.~\\

From (\ref{SUSYf4}), maximally supersymmetric  configuration, preserving
all the $\cN=4$ supersymmetries, exists
only if $\mass_{1}=0$. It corresponds to the static fuzzy sphere
\be
\ba{ll}
D_{t}X_{a}=0\,,~~~~&~~~~
\left[X_{a},X_{b}\right]=i\textstyle{\frac{1}{3}}\mass_{2}\,\epsilon_{abc}X^{c}\,.
\ea
\ee
Physically, when $\mu_{1}\neq 0$, the harmonic potential becomes
too steep to support the fuzzy sphere favored by  the Myers term.
It is worth to note that the Lagrangian (\ref{FINAL4}) can be rewritten such that the
fuzzy sphere structure is manifest in the potential
\be
\ba{ll}
\cL^{\4D}_{\typeI}=&
\tr\!\left[\half D_{t}X^{a}D_{t}X_{a}+\quarter
\Big([X^{a},X^{b}]-i\textstyle{\frac{1}{3}}\mass_{2}\,\epsilon_{abc}X^{c}\Big)^{2}
-\textstyle{\frac{1}{18}}\mu_{1}^{2}\,X^{a}X_{a}\right]\\
{}&{}\\
{}&+\,\tr\Big[-i\half\bar{\psi}\Gamma^{t}D_{t}\psi
-\half\bar{\psi}\Gamma^{a}[X_{a},\psi]
+i\textstyle{\frac{1}{4}}
\bar{\psi}\left(\mass_{1}+\mass_{2}\Gamma^{123}\right)\psi\Big]\,.
\ea
\label{FINAL4re}
\ee
~~\\

Finally we note an interesting property at  $\mass_{1}=0$.
In this case, a time dependent field redefinition of the fermion
\be
\psi~\longrightarrow~\dis{e^{\frac{1}{2} t (\mass_{3}-\mass_{2})\Gamma^{t123}}}\psi\,,
\ee
can change the mass of the fermion arbitrarily:
$\half \mass_{2}\rightarrow\half\mass_{3}$, without changing the mass of the bosons.
The resulting matrix model reads
\be
\ba{ll}
\left.\cL^{\4D}_{\typeI}\right|_{\mass_{1}=0}=&
\tr\!\left[\half D_{t}X^{a}D_{t}X_{a}+\quarter
\Big([X^{a},X^{b}]-i\textstyle{\frac{1}{3}}\mass_{2}\,\epsilon_{abc}X^{c}\Big)^{2}
\right]\\
{}&{}\\
{}&+\,\tr\Big[-i\half\bar{\psi}\Gamma^{t}D_{t}\psi
-\half\bar{\psi}\Gamma^{a}[X_{a},\psi]
+i\textstyle{\frac{1}{4}}\mass_{3}\bar{\psi}\Gamma^{123}\psi\Big]\,.
\ea
\label{FINAL4re0}
\ee
The mass parameter $\mass_{3}$ is fictitious,
since   different values can be mapped to one another  by the field redefinition.
The supersymmetry transformations are
\be
\ba{l}
\delta A_{0}=-i\bar{\psi}\Gamma_{t}\varepsilon(t)\,,~~~~~~~
\delta X_{a}=-i\bar{\psi}\Gamma_{a}\varepsilon(t)\,,\\
{}\\
\delta\psi=\Big(\!\!-\Gamma^{ta}D_{t}X_{a}+i\half[X_{a},X_{b}]\Gamma^{ab}
+\textstyle{\frac{1}{3}}\mass_{2} X\Gamma^{123} \Big)\varepsilon(t)\,,
\ea
\label{SUSYf4re0}
\ee
where
\be
\varepsilon(t)=\dis{e^{{\frac{1}{6}}
t\left(2\mass_{2}-3\mass_{3}\right)\Gamma^{t123}}\varepsilon(0)\,.}
\ee
The corresponding   supersymmetry algebra is now of the form
\be
\ba{lll}
{}\left[H,Q\right]=i\textstyle{\frac{1}{6}}
\left(2\mu_{2}-3\mass_{3}\right)\Gamma^{t123}Q\,,~~~~&~~~~
{}\left[R,Q\right]=i\Gamma^{t123}Q\,,~~~~&~~~~\left[H,R\right]=0\,,\\
{}&{}\\
\multicolumn{3}{c}{
{\left\{Q,\bar{Q}\right\}=
2\Big(\Gamma^{t}\!\left(
H+\half(\mass_{3}-\mass_{2})R\right)
-\textstyle{\frac{1}{6}}\mass_{2}\epsilon_{abc}\Gamma^{a}M^{bc}\Big)\,,}}
\ea
\ee
where, compared to (\ref{fTISA}), a new generator
$R=R^{\dagger}=i\half\bar{\psi}\Gamma^{123}\psi$  appears which corresponds to
the following  Noether symmetry of the Lagrangian
\be
\ba{lll}
\delta A_{0}=0\,,~~~~&~~~~\delta X_{0}=0\,,~~~~&~~~~
\delta \psi=\Gamma^{t123}\psi\,.
\ea
\ee
The superalgebra  is then    a central extension of $\su(2|1)$.
\subsection{Deformation $\SO(3)\rightarrow\SO(2)$,
$\mbox{Clifford}_{4}({\mathbf{R}})$ superalgebra : type II }
The other  deformation, type II,  of the $\cN=4$  super Yang-Mills quantum mechanics is
\be
\ba{ll}
\cL^{\4D}_{\typeII}\!=&\tr\Big(\half D_{t}X^{a}D_{t}X_{a}
+\quarter [X^{a},X^{b}]^{2}
-i\half\bar{\psi}\Gamma^{t}D_{t}\psi
-\half\bar{\psi}\Gamma^{a}[X_{a},\psi]\Big)\\
{}&{}\\
{}&+\tr\Big(
i\textstyle{\frac{1}{8}}\mass\bar{\psi}\Gamma^{t12}\psi
-\textstyle{\frac{1}{72}}\mu^{2}
\left(X_{1}^{2}+X_{2}^{2}+4X_{3}^{2}\right)\Big)\,.
\ea
\label{FINAL42}
\ee
The supersymmetry transformations are
\be
\ba{l}
\delta A_{0}=-i\bar{\psi}\Gamma_{t}\varepsilon(t)\,,~~~~~~~
\delta X_{a}=-i\bar{\psi}\Gamma_{a}\varepsilon(t)\,,\\
{}\\
\delta\psi=\Big(-\Gamma^{ta}D_{t}X_{a}+i\half[X_{a},X_{b}]\Gamma^{ab}
-\textstyle{\frac{1}{4}}\mass \Gamma^{t12}X
+\textstyle{\frac{1}{12}}\mass X\Gamma^{t12} \Big)\varepsilon(t)\,,
\ea
\label{SUSYf4prime}
\ee
where
\be
\varepsilon(t)=\dis{e^{{-\frac{1}{12}}
t\mass\,\Gamma^{12}}
\varepsilon(0)\,.}
\ee
~\\

The supersymmetry algebra reads
\be
\ba{ll}
{}\left[H,Q\right]=i\textstyle{\frac{1}{12}}\mass\Gamma_{12}Q\,,~~~~&~~~~
{}\left\{Q,\bar{Q}\right\}=
2\Gamma^{t}\left(H-\textstyle{\frac{1}{6}}\mass M_{12}\right)\,.
\ea
\ee
Similar to  (\ref{central}), (\ref{unitaryb}),
$H-\textstyle{\frac{1}{6}}\mass M_{12}$ is central and positive semi-definite.
The corresponding superalgebra is then $\mbox{Clifford}_{4}({\mathbf{R}})$.\\

There exists a nontrivial  $\cN=4$ supersymmetric configuration  corresponding
to a circular motion of the $D$-particles,
\be
\ba{lll}
X_{1}=R\cos\left(\textstyle{\frac{1}{6}}t\mass\right)1\,,~~~~&~~~~
X_{2}=R\sin\left(\textstyle{\frac{1}{6}}t\mass\right)1\,,~~~~&~~~~X_{3}=0\,.
\ea
\ee
This saturates  the unitary bound
$H=\textstyle{\frac{1}{6}}\mass M_{12}=\left(\frac{1}{6}\mass R\right)^{2}$.\\
~\\
~\\
\section{$\cN=2$ super Yang-Mills quantum mechanics\label{3DSEC}}
The three-dimensional Minkowskian spacetime of the metric $\eta=\diag(-++)$
admits a  Majorana spinor, and  all the formulae in Section~\ref{4DSYM} for
the four-dimensional super Yang-Mills  can be freely  adopted for the
analysis on the three-dimensional super Yang-Mills. One only needs to note
that the gamma matrices are now $2\times 2$ and $\Gamma^{t12}=1$, such that
there are two real supersymmetries. \\


After the  dimensional reduction of the minimal  super Yang-Mills in three-dimensions,
we obtain a $\cN=2$ supersymmetric matrix model with
two Hermitian matrices  $X^{a}$, $a=1,2$,
\be
\cL^{\3D}_{0}=\tr\!\left(\half D_{t}X^{a}D_{t}X_{a}+\quarter [X^{a},X^{b}]^{2}
-i\half\bar{\psi}\Gamma^{t}D_{t}\psi
-\half\bar{\psi}\Gamma^{a}[X_{a},\psi]\right)\,.
\label{L03}
\ee
The most general mass term we may add  for the fermion,  which is compatible
with the Majorana condition, is
\be
i\textstyle{\frac{1}{8}}\mu\,\tr\!\left(\bar{\psi}\psi\right)\,.
\ee
On the other hand, there is no   supersymmetric counter term for the bosons which are linear in $\mu$,
up to the field redefinition (\ref{rotating}).
Obviously   the  Myers term can not exist with two spatial directions.
As we show in Appendix \ref{derN=2},
the supersymmetric completion of the above mass term is unique. The resulting
massive $\cN=2$  super Yang-Mills quantum mechanics  reads
\be
\ba{ll}
\cL^{\3D}_{\Mass}=&\tr\Big(\half D_{t}X^{a}D_{t}X_{a}+\half [X_{1},X_{2}]^{2}
-i\half\bar{\psi}\Gamma^{t}D_{t}\psi
-\half\bar{\psi}\Gamma^{a}[X_{a},\psi]\Big)\\
{}&{}\\
{}&+\tr\Big(i\textstyle{\frac{1}{8}}\mass\bar{\psi}\psi
-\textstyle{\frac{1}{72}}\mass^{2}\left(X_{1}^{2}+X_{2}^{2}\right)\Big)\,.
\ea
\label{FINAL3}
\ee
The supersymmetry transformations are
\be
\ba{l}
\delta A_{0}=-i\bar{\psi}\Gamma_{t}\varepsilon(t)\,,~~~~~~~
\delta X_{a}=-i\bar{\psi}\Gamma_{a}\varepsilon(t)\,,\\
{}\\
\delta\psi=\Big(-\Gamma^{ta}D_{t}X_{a}+i[X_{1},X_{2}]\Gamma^{12}
-\textstyle{\frac{1}{6}}\mass\Gamma^{a}X_{a}\Big)\varepsilon(t)\,,
\ea
\label{SUSYf3}
\ee
where
\be
\varepsilon(t)=\dis{e^{{\frac{1}{12}}t\mass\Gamma^{t}}
\varepsilon(0)\,.}
\ee
{}\\

With the Majorana supercharge $\bar{Q}=Q^{T}C=Q^{\dagger}\Gamma^{t}$,
the supersymmetry algebra of  the
$\cN=2$ massive super Yang-Mills quantum mechanics
reads
\be
\ba{ll}
{}\left[H,Q\right]=-i\textstyle{\frac{1}{12}}\mass\Gamma^{t}Q\,,~~~~&~~~~
{}\left\{Q,\bar{Q}\right\}=
2\Gamma^{t}\left(H-\textstyle{\frac{1}{6}}\mass M_{12}\right)\,.
\ea
\ee
As before (\ref{central}), (\ref{unitaryb}),
$H-\textstyle{\frac{1}{6}}\mass M_{12}$ is central and positive semi-definite.
The corresponding superalgebra is $\mbox{Clifford}_{2}({\mathbf{R}})$. \\

The BPS state preserving all the two supersymmetries corresponds to a circular motion
\be
\ba{ll}
X_{1}=R\cos\left(\textstyle{\frac{1}{6}}t\mass\right)1\,,~~~~&~~~~
X_{2}=R\sin\left(\textstyle{\frac{1}{6}}t\mass\right)1\,,
\ea
\label{BPS2}
\ee
which  saturates the unitary  bound
$H=\textstyle{\frac{1}{6}}\mass M_{12}=\left(\frac{1}{6}\mass R\right)^{2}$.

\section{$\cN={1+1}$ super Yang-Mills quantum mechanics\label{2DSEC}}
In two-dimensional  Minkowskian spacetime,  the Majorana-Weyl
spinor has only one real component, $\psi=\psi^{\dagger}$. Accordingly
the minimal super Yang-Mills field theory has   one real supersymmetry.
However,  it turns out that upon  dimensional reduction the number of supersymmetries is
doubled~\cite{Park:2005pz},  and hence our notation $\cN={1+1}$.
The most general supersymmetric deformation was obtained in \cite{Park:2005pz}
and shown  to allow for two arbitrary time dependent functions $\Lambda(t)$, $\rho(t)$,
\be
\displaystyle{
{\cal L}_{\Mass}^{\2D}=\tr\Big[\half\left(D_{t}X\right)^{2}+i\half\psi D_{t}\psi
+X\psi\psi +\half {\Lambda}(t)X^{2}+\rho(t)X\Big]\,.}
\label{FINAL2}
\ee
Here our convention is not to use  any gamma matrices so that $X$ is just
a Hermitian matrix.\footnote{Note that
in   the previous   sections,  our notation  was ``$X:=\Gamma^{a}X_{a}$".}  \\

The \textit{two} dynamical  supersymmetries,
which we distinguish by $+$, $-$ indices,  are
\be
\ba{ll}
\delta_{\pm} \A=\delta_{\pm} X=if_{\pm}(t)\psi\varepsilon_{\pm}\,, ~~~~~&~~~~~
\delta_{\pm}\psi=\Big(f_{\pm}(t)D_{t}X-\dot{f}_{\pm}(t)X-\kappa_{\pm}(t)
1\Big)\varepsilon_{\pm}\,.
\ea
\label{SUSYf2}
\ee
Here  $\varepsilon_{+}$, $\varepsilon_{-}$ are two real supersymmetry parameters;
$f_{+}(t)$, $f_{-}(t)$    are  two different solutions of the following
second order differential equation
\be
\dis{\ddot{f}_{\pm}(t)=f_{\pm}(t){\Lambda}(t)~;}
\ee
and $\kappa_{+}(t)$, $\kappa_{-}(t)$ are given by
\be
\dis{{\kappa}_{\pm}(t):=\int^{t}_{t_{0}}
\!{\rm d}t^{\prime}\,\rho(t^{\prime})f_{\pm}(t^{\prime})\,.}
\ee
These two dynamical supersymmetries further reveal three bosonic symmetries
which we denote by $~\delta_{\scriptscriptstyle{++}}$,
$~\delta_{\scriptscriptstyle{{--}}}$,
$~\delta_{\scriptscriptstyle{{\{+,-\}}}}$,
{~}in order to indicate their origins  as the anti-commutator of two supersymmetries:
\be
\ba{l}
\delta_{\scriptscriptstyle{++}} \A=\delta_{\scriptscriptstyle{++}} X=f_{+}\left(f_{+}
D_{t}X-\dot{f}_{+}X-\kappa_{+}1\right)\,,~~~~~~~
\delta_{\scriptscriptstyle{++}}\psi=0\,,\\
{}\\
\delta_{\scriptscriptstyle{--}} \A=\delta_{\scriptscriptstyle{--}} X=f_{-}\left(f_{-}
D_{t}X-\dot{f}_{-}X-\kappa_{-}1\right)\,,~~~~~~~
\delta_{\scriptscriptstyle{--}}\psi=0\,,\\
{}\\
\delta_{\scriptscriptstyle{{\{+,-\}}}} \A=\delta_{\scriptscriptstyle{{\{+,-\}}}} X=2f_{+}f_{-}D_{t}X-\left(f_{+}\dot{f}_{-}+f_{-}\dot{f}_{+}\right)X-
\left(f_{+}\kappa_{-}+f_{-}\kappa_{+}\right)1\,,~~
\delta_{\scriptscriptstyle{{\{+,-\}}}}\psi=0\,.
\ea
\label{hidden}
\ee
These bosonic symmetries form $\mbox{sp}(2,{\mathbf{R}})\equiv\so(1,2)$ Lie algebra,
and  with the two supersymmetries they form $\mbox{osp}(1|2,{\mathbf{R}})$
super Lie algebra.\\

From the supersymmetry transformations of the fermion, the BPS equations are
\be
f_{\pm}(t)D_{t}X=\dot{f}_{\pm}(t)X+\kappa_{\pm}(t) 1\,.
\label{BPSeq}
\ee
A generic BPS configuration then decomposes
into the traceless and  $\mathbf{u}(1)$ parts,
\be
\ba{lll}
X(t)=f_{+}(t){\cX}+h_{+}(t)1~~~~~&~~~\mbox{or}~~~&~~~~~
X(t)=f_{-}(t){\cX}+h_{-}(t)1\,.
\label{BPSconf}
\ea
\ee
Here ${\cX}$ is an arbitrary
traceless constant matrix, and $h_{\pm}(t)$ are
the solutions of the first order differential equation
$f_{\pm}\dot{h}_{\pm}=\dot{f}_{\pm}h_{\pm}+\kappa_{\pm}$
corresponding to  the center of mass $N^{-1}\tr X(t)=h_{\pm}(t)$.
Since $f_{+}(t)\neq f_{-}(t)$, the BPS state preserves only one supersymmetry. \\
~\\
~\\

\section{Comments\label{COMMENTSEC}}
Alternative to our approach \textit{i.e.~}looking for the supersymmetric completions
of  the  Yang-Mills quantum mechanics after  adding mass terms for fermions,
one can consider the
light-cone formulation of  the supersymmetric Nambu-Goto action
in any supersymmetric background and
try to obtain a corresponding
 supersymmetric matrix model,  as was done for the
eleven-dimensional background~\cite{deWit,deWit:1998tk}. \\


Conversely, from our resulting massive super Yang-Mills quantum mechanics, utilizing
the light-cone formulation (\ref{GeneralMM})  for the generic
background (\ref{Mmetric0}),
it is straightforward to deduce the corresponding supersymmetric  background for each
massive super Yang-Mills quantum mechanics.
For example, while the background for the  BMN matrix model~(\ref{BMN}) is
 the eleven-dimensional pp-wave background~(\ref{ppw}),  for
 the $\cN=4$ type II massive super Yang-Mills quantum mechanics,
 the relevant background is
\begin{equation}
ds^{2}=2dx^{+}dx^{-}-\textstyle{\frac{1}{36}}\mu^{2}\Big(
x_{1}^{2}+x_{2}^{2}+4x_{3}^{2}\Big)dx^{+}dx^{+}+
\displaystyle{\sum_{a=1}^{3}}\,dx^{a}dx^{a}\,.
\end{equation}
This  background preserves 8 supersymmetries~\cite{Gauntlett:2002nw}
which match the sum of the
 dynamical and kinematical  supersymmetries in
 $\cN=4$ type II massive super Yang-Mills quantum mechanics.\\
 ~\\
 ~\\

~\\
\newline
\begin{center}
\large{\textbf{Acknowledgments}}
\end{center}
Jeong-Hyuck Park wishes to thank  Leonard Susskind,  Sergei Kuzenko for
helpful comments  and  Korea Institute for Advanced Study for the hospitality
during a workshop on string theory,  12-16 June.
Nakwoo Kim is grateful to Institut des Hautes  \'{E}tudes Scientifiques  for
the hospitality  during a visit in summer
2005,  where this work was initiated.  The research of Nakwoo Kim is supported by
the Science Research Center Program of the Korea Science and Engineering Foundation
(KOSEF) through the Center for Quantum Spacetime (CQUeST) of Sogang University
with grant number R11-2005-021, and by the Basic Research Program of KOSEF with
grant No. R01-2004-000-10651-0. The research of Jeong-Hyuck Park is supported by
the Alexander von Humboldt foundation.
\newpage
\appendix
\section{Derivation of the mass deformations\label{DERIVATION}}

\subsection{$\cN=8$ super Yang-Mills quantum mechanics\label{derivation6}}
After the  dimensional reduction to  time $t$, the six-dimensional
super Yang-Mills  gives a
supersymmetric matrix model Lagrangian $\cL^{\6D}_{0}$ containing
five Hermitian matrices   $X^{a}$, $a=1,2,3,4,5$,
\be
\cL^{\6D}_{0}=\tr\!\left(\half D_{t}X^{a}D_{t}X_{a}+\quarter [X^{a},X^{b}]^{2}
-i\half\bar{\psi}^{i}\Gamma^{t}D_{t}\psi_{i}
-\half\bar{\psi}^{i}\Gamma^{a}[X_{a},\psi_{i}]\right)\,.
\label{L06A}
\ee
The mass dimensions are $1$ for the bosons and $\frac{3}{2}$ for the fermions so that
the Lagrangian has mass dimension $4$.\\

Dropping the $\su(2)$ indices of the $\su(2)$ Majorana-Weyl spinor,
\textit{i.e.~}$\psi\equiv\psi_{1}$, we can rewrite the Lagrangian as
\be
\cL^{\6D}_{0}=\tr\!\left(\half D_{t}X^{a}D_{t}X_{a}+\quarter [X^{a},X^{b}]^{2}
-i\bar{\psi}\Gamma^{t}D_{t}\psi
-\bar{\psi}\Gamma^{a}[X_{a},\psi]\right)\,.
\label{L06p}
\ee
We look for the mass deformation of the above matrix model
\be
\cL^{\6D}_{\Mass}=\cL^{\6D}_{0}+\mu\cL^{\6D}_{1}+\mu^{2}\cL^{\6D}_{2}+\cdots\,,
\label{GEXPF}
\ee
where $\mu$ is a constant mass parameter we introduce.
Accordingly, $\cL^{\6D}_{1}$ has  mass dimension three, and hence it should
take the form:\footnote{In (\ref{L1}), the chirality of the fermion has been taken
into account, so that $\Gamma^{tab}$ is absent in the expansion of $M$.}
\be
\ba{l}
\dis{\cL^{\6D}_{1}=\tr\!\left(\bar{\psi}M
\psi+\textstyle{\frac{1}{\,3!}}S_{abc}X^{a}X^{b}X^{c}
+J_{ab}X^{a}D_{t}X^{b}\right)\,,}\\
{}\\
\dis{
M=M_{t}\Gamma^{t}+M_{a}\Gamma^{a}-i\textstyle{\frac{1}{\,3!}}M_{abc}\Gamma^{abc}\,,}
\label{L1}
\ea
\ee
where $M_{t}$, $M_{a}$, $M_{abc}=M_{[abc]}$,
$S_{abc}$, $J_{ab}$ are dimensionless and may depend on time.
On the other hand, $\cL^{\6D}_{2}$ has mass dimension two. Hence it should be purely
bosonic and, in fact, quadratic in $X$~:
\be
\dis{\cL^{\6D}_{2}=-\tr\!\left(\half S_{(ab)}X^{a}X^{b}\right)\,.}
\ee
It is clear that the expansion (\ref{GEXPF}) terminates at the second order in $\mu$.\\

Without loss of generality we may set $J_{ab}=-J_{ba}$, since
\be
\dis{\tr\!\left(J_{ab}X^{a}D_{t}X^{b}\right)=\tr\!\left(J_{[ab]}X^{a}D_{t}X^{b}\right)
+\frac{\rmd~}{\rmd t}\tr\!\left(\half J_{(ab)}X^{a}X^{b}\right)
-\tr\!\left(\half\dot{J}_{(ab)}X^{a}X^{b}\right)\,,}
\label{Jset01}
\ee
and  the time derivative of
$J_{ab}$  corresponds to $\cL^{\6D}_{2}$. A time dependent
$\SO(5)$ rotation of the dynamical variables
$(X^{a},\psi)~\rightarrow~(L^{a}{}_{b}X^{b},\hat{L}\psi)$ with
\be
\ba{llll}
L^{T}=L^{-1}\,,~~~&~~~\hat{L}^{\dagger}=\hat{L}^{-1}\,,
~~~&~~~\hat{L}\Gamma_{a}\hat{L}^{-1}
=\Gamma_{b}L^{b}{}_{a}\,,~~~&~~~\hat{L}\Gamma_{t}=\Gamma_{t}\hat{L}\,,
\ea
\label{rotating}
\ee
leaves the above generic form of the Lagrangian (\ref{GEXPF}) invariant.
Among others, it induces a shift of $J_{ab}$,
\be
J_{ab}~\longrightarrow~J_{ab}-2\!\left(\!L^{T}\dot{L}\right)_{ab}\,.
\label{Jset02}
\ee
Hence, by solving $\dot{L}=\half LJ$ for a given $J$, we can eliminate $J_{ab}$
~\cite{Hyun:2002fk,Hyun:2003se}.
Therefore without loss of generality, we may set $J_{ab}=0$ in (\ref{L1}). Similarly
by a time dependent $\mbox{U}(1)$ phase transformation of the fermion we can set
$M_{t}=0$ as well,  such that
\be
\dis{M=M_{a}\Gamma^{a}-i\textstyle{\frac{1}{6}}M_{abc}\Gamma^{abc}\,.}
\ee

Now it is useful to note that
\be
\ba{ll}
\Gamma^{t}M\Gamma^{t}=M\,,~~&~~~
\Gamma^{a}M\Gamma^{b}-\Gamma^{b}M\Gamma^{a}+M\Gamma^{ab}
=4iM^{abc}\Gamma_{c}-i\Gamma^{ab}M\,,\\
{}&{}\\
\left[M,\Gamma^{12345}\right]=0\,,~~&~~~\dis{
M^{2}=M_{a}M^{a}+\textstyle{\frac{1}{6}}M_{abc}M^{abc}-iM_{abc}M^{c}\Gamma^{ab}
-\textstyle{\frac{1}{4}}M_{abe}M_{cd}{}^{e}\Gamma^{abcd}\,.}
\ea
\label{MGG}
\ee
The supersymmetry transformations for $\cL_{0}^{\6D}$  (\ref{susy0})
should get modified. However, in order to benefit from  the  Fierz identity,
(\ref{Fierz}),  we keep the transformation of the bosons as before,
but allow the supersymmetry parameter to be time dependent so that
\be
\ba{l}
\delta A_{0}=\bar{\psi}\Gamma_{0}\varepsilon(t)+
\bar{\varepsilon}(t)\Gamma_{0}\psi\,,
~~~~~~~\delta X_{a}=\bar{\psi}\Gamma_{a}\varepsilon(t)+
\bar{\varepsilon}(t)\Gamma_{a}\psi\,,\\
{}\\
\delta\psi=\left(
-iD_{t}X_{a}\Gamma^{ta}-\half\left[X_{a},X_{b}\right]\Gamma^{ab}+
\mass\Delta\right)\varepsilon(t)\,,
\ea
\label{susymuA}
\ee
where $\Delta$  is  Lie algebra valued, depends on $X^{a}$ and has
mass dimension one.
Its explicit form is  to be determined.
Although we allow $\varepsilon(t)$ to be  time dependent,
it can not be Lie algebra valued. Explicitly,   with a constant and chiral
supersymmetry parameter $\hat{\varepsilon}$, we set
\be
\ba{lll}
\varepsilon:=G(t)\hat{\varepsilon}\,,~~~~&~~~~
\partial_{t}\varepsilon(t)=\mass\Pi(t)\varepsilon(t)\,,
~~~~&~~~~\mass\Pi(t):=\partial_{t}G(t)G(t)^{-1}\,.
\ea
\ee
Accordingly under the above transformations we obtain
\be
\ba{l}
\delta\cL^{\6D}_{0}\simeq\mu
\tr\!\Big[\bar{\psi}\Gamma^{t}\left(-iD_{t}\Delta
+\Gamma^{ta}[X_{a},\Delta]-D_{t}X_{a}\Gamma^{ta}\Pi+i\half [X_{a},X_{b}]\Gamma^{ab}\Pi
-i\mu\Delta\Pi\right)\varepsilon\Big]+\cc\,,\\
{}\\
\delta\tr\left[\bar{\psi}M\psi\right]=
\tr\!\left[\bar{\psi}M\left(
-iD_{t}X_{a}\Gamma^{ta}-\half\left[X_{a},X_{b}\right]\Gamma^{ab}+
\mass\Delta\right)
\varepsilon\right]+\cc\,,\\
{}\\
\delta\tr\!\left[\textstyle{\frac{1}{\,3!}}S_{abc}X^{a}X^{b}X^{c}\right]=
\tr\!\left[\half\bar{\psi}S_{abc}X^{a}X^{b}\Gamma^{c}\varepsilon\right]+\cc\,,
\ea
\ee
where $\cc$ denotes the complex conjugate containing
$\bar{\varepsilon}$, and `$\simeq$' for $\delta\cL_{0}^{\6D}$ indicates that
total derivative terms are neglected. From these,
we can read off the expression for $\delta\cL^{\6D}_{\Mass}$ (\ref{GEXPF}).
Especially the terms involving $D_{t}X^{a}$ read
\be
\delta\cL^{\6D}_{\Mass}~\Longrightarrow~
\mu\tr\!\left[\bar{\psi}D_{t}\left(-i\Gamma^{t}\Delta+X_{a}\Gamma^{a}\Pi-
iM\Gamma^{t}X_{a}\Gamma^{a}\right)\varepsilon\right]+\cc\,,
\ee
which must vanish by itself. Hence we have
\be
\Delta\left(1+\Gamma^{(7)}\right)=\left(MX-iX\Gamma^{t}
\Pi\right)
\left(1+\Gamma^{(7)}\right)\,,
\ee
where we set $X:=X_{a}\Gamma^{a}$.
Further, since $\Delta\left(1-\Gamma^{(7)}\right)$ is irrelevant for the chiral
 spinor, we may simply set
\be
\Delta=MX-iX\Gamma^{t}\Pi\,.
\label{Delta6}
\ee
Now using (\ref{MGG}), it is straightforward to see
\be
\ba{l}
\dis{\delta\left[\cL_{0}^{\6D}+\mu\tr\Big(\bar{\psi}\left(
M_{a}\Gamma^{a}-i\textstyle{\frac{1}{\,3!}}M_{abc}\Gamma^{abc}\right)
\psi+i\textstyle{\frac{4}{3}}
M_{abc}X^{a}X^{b}X^{c}\Big)\right]}\\
{}\\
=\half\mu\tr\!\Big[\bar{\psi}[X_{a},X_{b}]\Gamma^{ab}
\left(M\!+3i\Gamma^{t}\Pi\right)\varepsilon\Big]
\!+\!\mu\tr\!\Big[\bar{\psi}\left(\mu M\Delta-i\mu\Gamma^{t}\Delta\Pi
-X\dot{\Pi}+i\dot{M}\Gamma^{t}X\right)\varepsilon\Big]\!+\cc
\ea
\label{semideltaL}
\ee
which shows
\be
\dis{\Pi=-i\textstyle{\frac{1}{3}}\Gamma^{t}M\,,}
\label{Pi6}
\ee
and fixes $\cL_{1}^{\6D}$ as
\be
S_{abc}=i8M_{abc}\,.
\label{S6}
\ee
Provided these,
(\ref{semideltaL}) further reduces to
\be
\ba{l}
\dis{\delta\left[\cL_{0}^{\6D}+\mu\tr\Big(\bar{\psi}\left(
M_{a}\Gamma^{a}-i\textstyle{\frac{1}{\,3!}}M_{abc}\Gamma^{abc}\right)
\psi+i\textstyle{\frac{4}{3}}
M_{abc}X^{a}X^{b}X^{c}\Big)\right]}\\
{}\\
=\mu^{2}\tr\!\Big[\bar{\psi}\left(M^{2}X+\textstyle{\frac{2}{3}}MXM+
\textstyle{\frac{1}{9}}XM^{2}-i\mu^{-1}\Gamma^{t}\left(\dot{M}X+
\textstyle{\frac{1}{3}}X\dot{M}\right)\right)\varepsilon\Big]\!+\cc\,,
\ea
\label{finalL}
\ee
and the right hand side  must be identified as the variation of
$-\mu^{2}\cL_{2}^{\6D}$ which is purely bosonic and quadratic in $X$.
Thus, the quantity inside the bracket between $\bar{\psi}$ and $\varepsilon$
must be a linear combination of  gamma matrices  up to the chiral projection. Hence,
\be
\dis{M^{2}X+\textstyle{\frac{2}{3}}MXM+
\textstyle{\frac{1}{9}}XM^{2}+i\mu^{-1}\left(\dot{M}X+
\textstyle{\frac{1}{3}}X\dot{M}\right)\Gamma^{12345}
=\Gamma^{b}S_{(ab)}\,,}
\ee
or equivalently  for  all $a=1,2,3,4,5$,
\be
\dis{M^{2}\Gamma_{a}+\textstyle{\frac{2}{3}}M\Gamma_{a}M+
\textstyle{\frac{1}{9}}\Gamma_{a}M^{2}+i\mu^{-1}\left(\dot{M}\Gamma_{a}+
\textstyle{\frac{1}{3}}\Gamma_{a}\dot{M}\right)\Gamma^{12345}
=S_{(ab)}\Gamma^{b}\,.}
\label{Msolve2}
\ee
Since $M$, $\Gamma^{a}$  are Hermitian and $S_{(ab)}$'s are  real, from
$S_{(ab)}\Gamma^{b}-\left(S_{(ab)}\Gamma^{b}\right)^{\dagger}=0$ we get
\be
\Gamma^{a}M^{2}-M^{2}\Gamma^{a}=i\textstyle{\frac{3}{2}}\mu^{-1}
\left(\Gamma^{a}\dot{M}+\dot{M}\Gamma^{a}\right)\Gamma^{12345}\,.
\ee
Contracting this with $\Gamma^{a}$ from the left and from the right separately, we
get
\be
\ba{l}
5M^{2}-\Gamma^{a}M^{2}\Gamma_{a}=
-4iM_{abc}M^{c}\Gamma^{ab}
-8\textstyle{\frac{1}{4}}M_{abe}M_{cd}{}^{e}\Gamma^{abcd}=0\,,\\
{}\\
5\dot{M}+\Gamma^{a}\dot{M}\Gamma_{a}=2\dot{M}_{a}\Gamma^{a}-i\dot{M}_{abc}\Gamma^{abc}
=0\,.
\ea
\ee
Thus, $M$ must be time independent $\dot{M}=0$ and satisfy
\be
\ba{ll}
M_{abe}M_{cd}{}^{e}\epsilon^{abcdf}=0\,,~~~~~&~~~~~M_{abc}M^{c}=0\,.
\ea
\label{twocons}
\ee
To solve these two constraints we set
\be
M_{abc}:=\half\epsilon_{abcde}M^{de}\,,
\ee
and take the canonical form for $M^{ab}$ utilizing the $\SO(5)$ rotation~:
\be
\dis{
M^{ab}=c_{1}\left(\ba{ccccc}
0&\cos\theta&0&0&~0\\
-\cos\theta&0&0&0&~0\\
0&0&0&\sin\theta&~0\\
0&0&-\sin\theta&0&~0\\
0&0&0&0&~0
\ea\right)^{\!ab}\,,}
\ee
where $c_{1}$ and $\theta$ are some constants.
The first constraint in (\ref{twocons}) implies $\epsilon_{abcde}M^{bc}M^{de}=0$ so that
\be
c_{1}\cos{\theta}\sin{\theta}=0\,.
\ee
Hence, without loss of generality, we get $\theta=0$. The second constraint
then shows $M_{3}=M_{4}=M_{5}=0$.  Using the $\SO(2)$ rotation we further set $
M_{a}=c_{2}\delta^{1}_{a}$, and hence all together
\be
M=c_{2}\Gamma^{1}-ic_{1}\Gamma^{345}\,.
\ee
Substituting this into (\ref{Msolve2}) we obtain the final constraint
\be
c_{1}c_{2}=0\,.
\ee
This implies that there are two distinct mass deformations of the $\cN=8$ super Yang-Mills
quantum mechanics: type I and type II.
The case  $c_{1}\neq 0$, $c_{2}=0$ corresponds to a deformation compatible with the
$\su(2)$ Majorana-Weyl spinors (type I), while the other deformation by
$c_{1}=0$, $c_{2}\neq 0$ is not so (type II).
The final results  on the mass deformations of
the $\cN=8$ super Yang-Mills quantum mechanics
are spelled out in (\ref{FINAL6}) for type I and (\ref{FINAL62}) for type II.

\subsection{$\cN=4$ super Yang-Mills quantum mechanics\label{derN=4}}
After the  dimensional reduction to the time, the four-dimensional
super Yang-Mills  gives a supersymmetric matrix model
$\cL^{\4D}_{0}$ containing  three Hermitian matrices  $X^{a}$, $a=1,2,3$,
\be
\cL^{\4D}_{0}=\tr\!\left(\half D_{t}X^{a}D_{t}X_{a}+\quarter [X^{a},X^{b}]^{2}
-i\half\bar{\psi}\Gamma^{t}D_{t}\psi
-\half\bar{\psi}\Gamma^{a}[X_{a},\psi]\right)\,.
\label{L04App}
\ee
Most general mass terms for fermions
 which are compatible with the Majorana condition are
\be
\mu\cL^{\4D}_{\psi}=-i\mu\tr\!\Big[\bar{\psi}
\left(c\Gamma^{123}+\Gamma^{t}H+r\cos\theta+r\sin\theta\Gamma^{t123}\right)
\psi\Big]\,,
\ee
where $H=\half\Gamma^{ab}H_{ab}$ and all the coefficients, $c,r$ are real
with $0\leq r$, $0\leq \theta<2\pi$. Under the chiral transformation
\be
\ba{ll}
\Big(\,\psi\,,~~\bar{\psi}\,\Big)
~~~\longrightarrow~~~\Big(\dis{e^{\phi\Gamma^{t123}}\psi}\,,
~~\bar{\psi}e^{\phi\Gamma^{t123}}\Big)\,,
~~~~&~~~~0\leq \phi<2\pi\,,
\ea
\label{chiralrotation}
\ee
$\cL_{0}^{\4D}$ is invariant, while the above mass terms for the fermions transform as
$\theta\rightarrow\theta+2\phi$. Thus, without loss of generality, we fix $\theta=0$ henceforth.

For the supersymmetric counter terms for the bosons   which are linear in $\mu$,
Myers term is the unique candidate
up to  field redefinitions, as discussed  in Section~\ref{derivation6},
\be
\mu\cL_{\Myers}=\mu\tr\Big(4i[X_{1},X_{2}]X_{3}\Big)\,.
\ee

Now letting the modified supersymmetry transformations be
\be
\ba{l}
\delta A_{0}=-i\bar{\psi}\Gamma_{t}\varepsilon(t)\,,~~~~~~
\delta X_{a}=-i\bar{\psi}\Gamma_{a}\varepsilon(t)\,,\\
{}\\
\delta\psi=\Big(-\Gamma^{ta}D_{t}X_{a}+i\half[X_{a},X_{b}]\Gamma^{ab}+\mu\Delta\Big)
\varepsilon(t)\,, \\
{}\\
\partial_{t}\varepsilon(t)=\mu\Pi\varepsilon(t)\,,
\ea
\label{susy4ans}
\ee
we consider the terms linear and quadratic in   $\mu$ appearing in  $\delta\left(\cL_{0}^{\4D}+\mu\cL^{\4D}_{\psi}+e\mu \cL_{\Myers}\right)$, where a real coefficient $e$ is introduced in order to allow   the case of the Myers term being absent. Without loss of generality we can take $e=0$ or $e=1$.  The linear terms containing $D_{t}$ read
\be
-i\mu\tr\!\left[\bar{\psi}\Big(D_{t}X\Pi+\Gamma^{t}D_{t}\Delta
-2(c\Gamma^{123}+\Gamma^{t}H+r)\Gamma^{t}D_{t}X\Big)\varepsilon\right]\,.
\ee
Requiring this to vanish, we get
\be
\Delta=2\left(r+\Gamma^{t}H-c\Gamma^{123}\right)X+\Gamma^{t}X\Pi\,.
\label{Delta4}
\ee
Consequently,  with $F=\half\Gamma^{ab}F_{ab}$ and using
\be
\ba{ll}
\Gamma^{a}D_{a}X=2F\,,~~~~&~~~~\Gamma^{a}HD_{a}X=F_{ab}H^{ab}\,1\,,
\ea
\ee
the remaining linear terms become
\be
-i\mu\tr\!\left[\bar{\psi}F\Big(
-3\Gamma^{t}\Pi+2r+2\Gamma^{t}H+2(2e-3c)\Gamma^{123}\Big)\varepsilon\right]\,.
\ee
Hence, to make this vanish for arbitrary $\psi$ and $\varepsilon$, we should set
\be
\Pi=\textstyle{\frac{1}{3}}\left(-2r\Gamma^{t}+(6c-4e)\Gamma^{t123}+2H\right)\,.
\ee
Substituting this into (\ref{Delta4}), we also obtain
\be
\Delta=\textstyle{\frac{4}{3}}\left(r-e\Gamma^{123}\right)X+2\Gamma^{t}HX+
\textstyle{\frac{2}{3}}\Gamma^{t}XH\,.
\ee

The quadratic terms are now
\be
-i\mu^{2}\tr\!\left[\bar{\psi}\Big(\Gamma^{t}\Delta\Pi+2(r+c\Gamma^{123}+\Gamma^{t}H)\Delta
\Big)\varepsilon\right]\,,
\ee
where explicitly,
\be
\ba{l}
\Gamma^{t}\Delta\Pi+2(r+c\Gamma^{123}+\Gamma^{t}H)\Delta\\
{}\\
=\textstyle{\frac{16}{9}}\Gamma^{t}\left(XH+3HX\right)\left(r-e\Gamma^{123}\right)
-\textstyle{\frac{8}{3}}HXH-4H^{2}X-\textstyle{\frac{4}{9}}XH^{2}\\
{}\\
~~\,+\textstyle{\frac{16}{9}}X\Big[e^{2}+r^{2}+3r(c-e)\Gamma^{123}\Big]\,.
\ea
\label{second4}
\ee
For the supersymmetry invariance, this quadratic terms must be cancelled by the variation of $\cL_{2}^{\4D}$ which is purely  bosonic. This  implies that  in  the  expansion of
(\ref{second4}) by the gamma matrix products
$\left\{1,\Gamma_{\mu}, \Gamma_{\mu\nu}, \Gamma_{\lambda\mu\nu}, \Gamma_{t123}\right\}$,
 only the linear order gamma matrices, $\Gamma_{a}$, $a=1,2,3$  must appear.
 In particular, $\Gamma_{t}$ is not also allowed, as the explicit appearance of
 the gauge field $A_{0}$, would break the gauge symmetry. From
\be
XH+3HX=2\epsilon_{abc}X^{a}H^{bc}\Gamma^{123}+2H_{ab}X^{b}\Gamma^{a}\,,
\ee
the term linear in $\Gamma^{t}$  is $\frac{32}{9}e\Gamma^{t}
\epsilon_{abc}X^{a}H^{bc}$. Since this  should vanish for arbitrary $X^{a}$,
we must require
\be
\ba{lll}
H=0~~:~~\mbox{type~I}\,,~~~~~~~&\mbox{or}&~~~~~~~e=0~~:~~\mbox{type~II}\,.
\ea
\ee
~\\

If $H=0$ (type I),   Eq.(\ref{second4}) gets simplified and we note $c=e$ or $r=0$.
However, when $r=0$,  a field redefinition  of the fermions which is given by
a time dependent chiral rotation
$\psi\rightarrow \exp(-2\mu \delta c\Gamma^{t123}t)\psi$, shifts
the constant $c$ only as $c\rightarrow c+\delta c$.  Thus, the  $r=0$ case can be
treated as a special case of $c=e$. \\

On the other hand, if $e=0$  and $H\neq 0$ (type II), we observe $r$ should vanish,
and hence, as explained above, we can safely set $c=0$. A canonical choice
 $H=\half\Gamma_{12}$ gives
\be
\displaystyle{\delta\left[\cL_{0}^{\4D}+\mu\cL^{\4D}_{\psi}-
\textstyle{\frac{2}{9}}\mu^{2}
\left(X_{1}^{2}+X_{2}^{2}+4X_{3}^{2}\right)\right]=\mbox{total~derivative}\,.}
\ee
This completes our analysis.
The final results on two different  mass deformations of $\cN=4$
super Yang-Mills quantum mechanics are spelled out in (\ref{FINAL4}) for type I and
(\ref{FINAL42}) for type II.


\subsection{$\cN=2$ super Yang-Mills quantum mechanics\label{derN=2}}
The three-dimensional Minkowskian spacetime of the metric
$\eta=\diag(-++)$, admits a Majorana spinor, and  all the formulae
in Section~\ref{4DSYM} for the four-dimensional super Yang-Mills  can be freely
adopted, with the understanding that the gamma matrices are now $2\times 2$, and
$\Gamma^{t12}=1$. There are two real supersymmetries. \\

After the  dimensional reduction to the time, the  super Yang-Mills in three-dimensions
gives a supersymmetric matrix model $\cL^{\3D}_{0}$ which is
essentially  of the same form as (\ref{L04}) but contains
two Hermitian matrices $X^{a}$, $a=1,2$.\\

The most general mass term for fermions which is compatible with the Majorana
condition reads
\be
\mu\cL^{\3D}_{\psi}=-i\half\mu\,\tr\!\left(\bar{\psi}\psi\right)\,.
\ee
On the other hand, there is no   supersymmetric counter term for the bosons which is linear in
$\mu$,  up to the field redefinition (\ref{rotating}). Clearly  the  Myers term does
not exist with two spatial directions.\\

Now we take  the modified supersymmetry transformations
to be as before (\ref{susy4ans}), and
 consider the terms linear and quadratic in   $\mu$ appearing in
$\delta\left(\cL_{0}^{\4D}+\mu\cL^{\3D}_{\psi}\right)$. The linear terms containing
$D_{t}$ read
\be
-i\mu\tr\!\left[\bar{\psi}\Big(D_{t}X\Pi+\Gamma^{t}D_{t}\Delta-\Gamma^{t}D_{t}X\Big)
\varepsilon\right]\,.
\ee
Requiring this to vanish, we get
\be
\Delta=X+\Gamma^{t}X\Pi\,.
\label{Delta3}
\ee
Consequently,  with $F=\half\Gamma^{ab}F_{ab}$ and using  $\Gamma^{a}D_{a}X=2F$
the remaining linear terms become
\be
-i\mu\tr\!\left[\bar{\psi}F\left(
1-3\Gamma^{t}\Pi\right)\varepsilon\right]\,.
\ee
Hence
\be
\Pi=-\textstyle{\frac{1}{3}}\Gamma^{t}\,.
\ee
Substituting this into (\ref{Delta4}), we also obtain
\be
\Delta=\textstyle{\frac{2}{3}}X\,.
\ee
This completes our analysis.
The final result on the mass deformation of the  $\cN=2$
super Yang-Mills quantum mechanics is spelled out in (\ref{FINAL3}).

\subsection{$\cN={1+1}$ super Yang-Mills quantum  mechanics\label{derN=1+1}}
The dimensional reduction of the minimal super Yang-Mills in two-dimensions  leads to
 the following supersymmetric  matrix model
\be
\cL_{0}=\tr\Big[\half D_{t}XD_{t}X+i\half\psi D_{t}\psi
+X\psi\psi\Big]\,.
\label{2D0}
\ee
The supersymmetry transformation $\delta_{\YM}$
descending from the two-dimensional  super Yang-Mills   is,
with a constant supersymmetry parameter $\varepsilon$,
\be
\ba{ll}
\delta_{\YM} \A=\delta_{\YM}X=i\psi\varepsilon\,, ~~~~~&~~~~~
\delta_{\YM}\psi=D_{t}X\varepsilon\,.
\ea
\label{susy20}
\ee
Now, following \cite{Park:2005pz},
we look for the generalization  of the above Lagrangian
as well as the supersymmetry transformations.
First of all, we note from
\be
\tr\Big[i\half\psi D_{t}\psi
+X\psi\psi\Big]=\tr\Big[i\half\psi\partial_{t}\psi+(X-\A)\psi\psi\Big]\,,
\ee
that, in order to cancel the possible   cubic term of  $\psi$  which may
arise from the transformation of   $(X-\A)$,  it is inevitable   to impose
$\delta \A=\delta X$. Hence,  introducing   a time      dependent function $f(t)$,
we set the generalized supersymmetry transformation to be
\be
\ba{ll}
\delta \A=\delta X=if(t)\psi\varepsilon\,, ~~~~~&~~~~~
\delta\psi=\Big(f(t)D_{t}X+\Delta\Big)\varepsilon\,,
\ea
\label{susy2ans}
\ee
where $\Delta$ is a bosonic quantity having  mass dimension $2$, whose
explicit form  is to be determined shortly.
After some straightforward manipulation,  we obtain
\be
\delta{\cL_{0}}=\tr\!\left[
i\psi\varepsilon\left(D_{t}\Big(\dot{f}X+\Delta\Big)-\ddot{f}X+i[X,\Delta]\right)
\right]~+~\partial_{t}\cK\,,
\ee
where the total  derivative term is given by
\be
\cK=\tr\!\left(D_{t}X\delta X-i\half\psi\delta\psi\right)\,.
\ee
Of course, the simplest case where  $f(t)=1$ and $\Delta=0$ reduces to the
supersymmetry of the original  two-dimensional  super Yang-Mills, (\ref{susy20}).
For the generic cases,   we are obliged to  set
\be
\Delta=-\dot{f}X-\kappa 1\,,
\label{Deltafix}
\ee
and obtain the following supersymmetry invariance
\be
\delta\left[{\cL_{0}}+
\tr\!\left(\half{(\ddot{f}/f)}X^{2}+(\dot{\kappa}/f)X\right)\right]=\partial_{t}\cK\,.
\ee
This essentially leads to the final result (\ref{FINAL2}) with the
supersymmetry enhancement:  $\cN=1\rightarrow\cN=1+1$.
This kind of  supersymmetry enhancement after the dimensional reduction can
 be   noted  elsewhere \textit{e.g.~}\cite{Hyun:2002fk,Hyun:2003se}.
 A physical reason for the enhancement is that
 the D-brane  which  the higher dimensional field theory describes  preserves only a
   fraction of the supersymmetries of the corresponding  $\cM$-theory.

\subsection{$\cN={16}$ super Yang-Mills quantum mechanics\label{derN=16}}
Here we show that for $\cN={16}$ super Yang-Mills quantum mechanics,
the BMN matrix model (\ref{BMN}) is the unique mass deformation.
For simplicity we employ the real and symmetric nine-dimensional gamma matrices,
$\Gamma^{a}=\Gamma^{a}{}^{\ast}=\Gamma^{a}{}^{T}$, $a=1,2,3,\cdots,9$ such that
$\Gamma^{123456789}=1$
and the sixteen component Majorana spinor is  real.
The undeformed super Yang-Mills quantum mechanics follows from
the dimensional reduction of the ten-dimensional super Yang-Mills,
\be
 \cL^{\10D}_{0}=\tr\Big(\half D_{t}X^{a}D_{t}X_{a}+\quarter [X^{a},X^{b}]^{2}
 +i\half\Psi^{T}D_{t}\Psi-\half\Psi^{T}\Gamma^{a}[X_{a},\Psi]\Big)\,.
\ee
In a similar fashion to the previous analysis, henceforth
we look for the mass deformation of the above matrix model:
\be
\cL^{\10D}_{\Mass}=\cL^{\10D}_{0}+\mu\cL^{\10D}_{1}+\mu^{2}\cL^{\10D}_{2}\,.
\label{GEXPF10}
\ee
In particular, $\cL^{\10D}_{1}$ has  mass dimension three, and hence it should
take the form
\be
\dis{\cL^{\10D}_{1}=\tr\!\left(\half i\Psi^{T}M\Psi
+\textstyle{\frac{1}{\,3!}}S_{abc}X^{a}X^{b}X^{c}
+J_{ab}X^{a}D_{t}X^{b}\right)\,,}
\ee
where $M$ is a real and anti-symmetric matrix, so that it can be expressed as
\be
\dis{M=\half M_{ab}\Gamma^{ab}+\textstyle{\frac{1}{\,3!}}M_{abc}\Gamma^{abc}\,.}
\ee
As discussed before through (\ref{Jset01}), (\ref{rotating}), (\ref{Jset02}),
without loss of generality we may set $J_{ab}=0$, and consider the following
supersymmetric transformation
\be
\ba{lll}
\delta A_{0}=i\Psi^{T}\varepsilon(t)\,,~~~~&~~~~
\delta X^{a}=i\Psi^{T}\Gamma^{a}\varepsilon(t)\,,~~~~&~~~~
\delta \Psi=\left(D_{t}X+F+\mass\Delta\right)\varepsilon(t)\,,
\ea
\ee
where we set $X:=X^{a}\Gamma_{a}$, $F:=-i\half\left[X^{a},X^{b}\right]\Gamma_{ab}$.
We further write $\partial_{t}\varepsilon(t)=\mass\Pi\varepsilon(t)$. \\

The transformation induces
\be
\ba{l}
\displaystyle{
\delta\cL^{\10D}_{0}=i\mass\tr\!\left[\Psi^{T}\Big(
D_{t}\Delta+D_{t}X\Pi+F\Pi+i\Gamma^{a}\left[X_{a},\Delta\right]
+\mass\Delta\Pi\Big)\right]\varepsilon(t)\,+\,\mbox{total~derivative}\,,}\\
{}\\
\displaystyle{
\delta\cL^{\10D}_{1}=i\tr\!\left[\Psi^{T}\Big(M\left(
D_{t}X+F+
\mass\Delta\right)+\textstyle{\frac{1}{2}}S_{abc}X^{a}X^{b}\Gamma^{c}\Big)
\right]\varepsilon(t)\,.}
\ea
\ee
Requiring $\delta\cL^{\10D}_{\Mass}$ to vanish,   in analogy to
(\ref{Delta6}), (\ref{Pi6}), (\ref{S6}),  we obtain sequently
\be
\ba{l}
\Delta=-X\Pi-MX\,,\\
{}\\
\Pi=\textstyle{\frac{1}{3}}M\,,\\
{}\\
S_{abc}=-8iM_{abc}\,.
\ea
\ee
Hence, we have
\be
\delta\left(\cL^{\10D}_{0}+\mass\cL^{\10D}_{1}\right)=
i\mass\tr\left[\Psi^{T}F_{ab}M_{cd}\Gamma^{abcd}\right]\varepsilon(t)
+\cO\left(\mass^{2}\right)+\mbox{total~derivative}\,.
\ee
Thus for the supersymmetry  invariance it is required to set $M_{ab}=0$, which is
essentially due to our gauge choice $J_{ab}=0$.
The remaining terms are second order in $\mass$ as
\be
\delta\left(\cL^{\10D}_{0}+\mass\cL^{\10D}_{1}\right)
=\dis{-i\mass^{2}\tr\!\left[\Psi^{T}X_{a}\cM^{a}\varepsilon(t)\right]
+\mbox{\,total~derivative}\,,}
\ee
where we set
\be
\ba{ll}
\cM^{a}&:=\mass^{-1}\!\left(\dot{M}\Gamma^{a}+\textstyle{\frac{1}{3}}
\Gamma^{a}\dot{M}\right)+M^{2}\Gamma^{a}+\textstyle{\frac{2}{3}}M\Gamma^{a}M
+\textstyle{\frac{1}{9}}\Gamma^{a}M^{2}\\
{}&{}\\
{}&\,=-\textstyle{\frac{2}{27}}M_{bcd}M^{bcd}\Gamma^{a}-
\textstyle{\frac{2}{3}}M^{abc}M_{bcd}\Gamma^{d}+
\textstyle{\frac{2}{3}}\mass^{-1}\dot{M}^{a}{}_{bc}\Gamma^{bc}
-\textstyle{\frac{8}{9}}M^{a}{}_{bc}M_{de}{}^{c}\Gamma^{bde}\\
{}&{}\\
{}&~~~-\textstyle{\frac{1}{9}}\mass^{-1}\dot{M}_{bcd}\Gamma^{abcd}
+{\frac{1}{9}}M_{bcd}M_{ef}{}^{d}\Gamma^{abcef}
+{\frac{1}{9}}M^{a}{}_{bc}M_{def}\Gamma^{bcdef}\,.
\ea
\ee
This must be linear in gamma matrices, and hence from the second order term
we see $M$ must be time independent $\dot{M}_{abc}=0$. Other higher order terms
give the following two constraints
\be
\ba{ll}
M^{ab}{}_{[c}M_{de]b}=0\,,~~~~&~~~~M^{a}{}_{[bc}M_{def]}=0\,.
\ea
\label{Mconst10}
\ee
Now we look for the most general solution of these constraints,
up to the $\SO(9)$ rotation.
Without loss of generality we set $M_{9ab}M_{9}{}^{ab}\neq 0$.
Using the $\SO(9)$ rotation, we further let among the components $M_{9ij}$,
$1\leq i,j\leq 8$, only $M_{129}$, $M_{349}$, $M_{569}$, $M_{789}$ may be
non-vanishing, while $M_{789}$ is strictly nonzero.
Now, considering the case $a=d=9$, $b=7$, $c=8$
for the latter constraint in (\ref{Mconst10}),
we see that all the components $M_{9ij}$ are vanishing except $M_{789}$.
Moreover, the case $a=9$ for the constraint shows that in fact all the components
$M_{abc}$ are vanishing except  $M_{789}$. In this case, the first constraint in
(\ref{Mconst10}) is automatically satisfied and the resulting mass deformation of the
$\cN=16$ super Yang-Mills quantum mechanics is uniquely given by the BMN matrix model
(\ref{BMN}). This completes our proof.


\newpage



\begin{thebibliography}{99}

\bibitem{Berenstein:2002jq}
D.~Berenstein, J.~M.~Maldacena and H.~Nastase,
JHEP {\bf 0204} (2002) 013 [hep-th/0202021].



\bibitem{Kim:2002cr}
  N.~Kim, K.~M.~Lee and P.~Yi,
  JHEP {\bf 0211} (2002) 009
  [arXiv:hep-th/0207264].



\bibitem{Park:2005pz}
  J.-H.~Park,
Nucl.\ Phys.\ B {\bf 745} (2006) 123 [arXiv:hep-th/0510070].



\bibitem{Myers:1999ps}
  R.~C.~Myers,
  JHEP {\bf 9912} (1999) 022
  [arXiv:hep-th/9910053].



\bibitem{Bonelli:2002mb}
  G.~Bonelli,
   ``Matrix strings in pp-wave backgrounds from deformed super Yang-Mills
   theory,''
  %
  JHEP {\bf 0208} (2002) 022
  [arXiv:hep-th/0205213].



\bibitem{Susskind:1967rg}
  L.~Susskind,
  Phys.\ Rev.\  {\bf 165} (1968) 1535.

\bibitem{Kogut:1972di}
  J.~B.~Kogut and L.~Susskind,
  Phys.\ Rept.\  {\bf 8} (1973) 75.



\bibitem{deWit}
B.~de Wit, J.~Hoppe and H.~Nicolai,
Nucl.\ Phys.\ B {\bf 305} (1988) 545.


\bibitem{Hoppe}
J.~Hoppe, ``Membranes and matrix models,'' [hep-th/0206192].


\bibitem{Banks:1996vh}
T.~Banks, W.~Fischler, S.~H.~Shenker and L.~Susskind,
Phys.\ Rev.\ D {\bf 55} (1997) 5112 [hep-th/9610043].


\bibitem{Susskind:1997cw}
L.~Susskind, ``Another conjecture about M(atrix) theory,'' [hep-th/9704080].


\bibitem{Sen}
A.~Sen,
Adv.\ Theor.\ Math.\ Phys.\  {\bf 2} (1998) 51 [hep-th/9709220].


\bibitem{Seiberg}
N.~Seiberg,
Phys.\ Rev.\ Lett.\  {\bf 79} (1997) 3577 [hep-th/9710009].




\bibitem{Figueroa-O'Farrill:2002ft}
  J.~Figueroa-O'Farrill and G.~Papadopoulos,
  JHEP {\bf 0303} (2003) 048
  [arXiv:hep-th/0211089].




\bibitem{Kowalski-Glikman}
J.~Kowalski-Glikman,
Phys.\ Lett.\ B {\bf 134} (1984) 194.

\bibitem{Figueroa-O'Farrill1}
J.~Figueroa-O'Farrill and G.~Papadopoulos,
JHEP {\bf 0108} (2001) 036 [hep-th/0105308].

\bibitem{FO2}
M.~Blau, J.~Figueroa-O'Farrill, C.~Hull and G.~Papadopoulos,
Class.\ Quant.\ Grav.\  {\bf 19} (2002) L87 [hep-th/0201081].



\bibitem{Dasgupta:2002hx}
K.~Dasgupta, M.~M.~Sheikh-Jabbari and M.~Van Raamsdonk,
JHEP {\bf 0205} (2002) 056 [hep-th/0205185].


\bibitem{Gauntlett:2002nw}
  J.~P.~Gauntlett, J.~B.~Gutowski, C.~M.~Hull, S.~Pakis and H.~S.~Reall,
  Class.\ Quant.\ Grav.\  {\bf 20} (2003) 4587
  [arXiv:hep-th/0209114].

\bibitem{Tod}
  K.~Tod,
  %
  Phys.\ Lett.\ B {\bf 121}, 241 (1983);\\
  K.~Tod,
  %
  Class.\ Quant.\ Grav.\  {\bf 12}, 1801 (1995).






\bibitem{Horava:2005tt}
  P.~Horava and C.~A.~Keeler,
  arXiv:hep-th/0508024.

\bibitem{Horava:2005wm}
  P.~Horava and C.~A.~Keeler,
  Nucl.\ Phys.\ B {\bf 745} (2006) 1
  [arXiv:hep-th/0512325].


\bibitem{Petkou:2005se}
  A.~C.~Petkou and G.~Siopsis,
  ``M-theory and the Gross-Neveu model in 2+1 dimensions,''
  arXiv:hep-th/0509143.




\bibitem{McGuigan3DM}
  M.~McGuigan, ``Three dimensional gravity and M-theory,''
  arXiv:hep-th/0312327;\\
  M.~McGuigan, ``Noncritical M-theory: Three dimensions,''
  arXiv:hep-th/0408041;\\
  M.~McGuigan, ``Cosmological constant seesaw in quantum cosmology,''
  arXiv:hep-th/0602112;\\
  M.~McGuigan, ``Cosmological constant seesaw in string/M-theory,''
  arXiv:hep-th/0604108.


\bibitem{Gomis:2005ce}
  J.~Gomis,
  JHEP {\bf 0510} (2005) 095
  [arXiv:hep-th/0508132].




\bibitem{Kim:2002tj}
  N.~Kim and J.~T.~Yee,
  Phys.\ Rev.\ D {\bf 67} (2003) 046004
  [arXiv:hep-th/0211029].


\bibitem{Aharony:1997th}
  O.~Aharony, M.~Berkooz, S.~Kachru, N.~Seiberg and E.~Silverstein,
  %
  Adv.\ Theor.\ Math.\ Phys.\  {\bf 1}, 148 (1998)
  [arXiv:hep-th/9707079].

\bibitem{Aharony:1997an}
  O.~Aharony, M.~Berkooz and N.~Seiberg,
  %
  Adv.\ Theor.\ Math.\ Phys.\  {\bf 2}, 119 (1998)
  [arXiv:hep-th/9712117].



\bibitem{Kim:2003rz}
  N.~Kim, T.~Klose and J.~Plefka,
  Nucl.\ Phys.\ B {\bf 671} (2003) 359
  [arXiv:hep-th/0306054].



\bibitem{Nicolai:1988ek}
  H.~Nicolai, E.~Sezgin and Y.~Tanii,
  Nucl.\ Phys.\ B {\bf 305} (1988) 483.


\bibitem{Ishiki:2006rt}
  G.~Ishiki, Y.~Takayama and A.~Tsuchiya,
  ``N = 4 SYM on R x S**3 and theories with 16 supercharges,''
  arXiv:hep-th/0605163.



\bibitem{Erdmenger:2006eh}
  J.~Erdmenger, J.-H.~Park and C.~Sochichiu,
  ``Matrix models from D-particle dynamics and the string / black hole
  transition,''
  arXiv:hep-th/0603243.

\bibitem{non-Abelian}
T.~Hagiwara, J.\ Phys.\ A {\bf 14}, 3059 (1981);\\
P.~C.~Argyres and C.~R.~Nappi, Nucl.\ Phys.\ B {\bf 330}, 151 (1990);\\
  A.~A.~Tseytlin,  Nucl.\ Phys.\ B {\bf 501}, 41 (1997) [hep-th/9701125];\\
J.-H.~Park,
  Phys.\ Lett.\ B {\bf 458} (1999) 471 [hep-th/9902081];\\
 E.~Serie, T.~Masson and R.~Kerner,
  Phys.\ Rev.\ D {\bf 68} (2003) 125003 [hep-th/0307105];\\
E.~Serie, T.~Masson and R.~Kerner,
  Phys.\ Rev.\ D {\bf 70} (2004) 067701
  [hep-th/0408012];\\
  E.~Serie,
\textit{Ph.D. thesis}  arXiv:math-ph/0512094.



\bibitem{Witten:1995im}
  E.~Witten,
  Nucl.\ Phys.\ B {\bf 460} (1996) 335
  [arXiv:hep-th/9510135].



\bibitem{Park:2002eu}
  J.-H.~Park,
  Phys.\ Lett.\ A {\bf 307} (2003) 183
  [arXiv:hep-th/0203017].

\bibitem{Nambu:1973qe}
  Y.~Nambu,
  %
  Phys.\ Rev.\ D {\bf 7} (1973) 2405.



\bibitem{Curtright}
  T.~Curtright and C.~K.~Zachos,
  %
  Phys.\ Rev.\ D {\bf 68} (2003) 085001
  [arXiv:hep-th/0212267];\\
  T.~Curtright and C.~K.~Zachos,
  %
  AIP Conf.\ Proc.\  {\bf 672} (2003) 165
  [arXiv:hep-th/0303088].



\bibitem{Sheikh-Jabbari:2004ik}
  M.~M.~Sheikh-Jabbari,
  JHEP {\bf 0409} (2004) 017  [arXiv:hep-th/0406214].

\bibitem{Park:2003ku}
  J.-H.~Park,
  JHEP {\bf 0309} (2003) 046
  [arXiv:hep-th/0307060].

\bibitem{Seiberg:1999vs}
  N.~Seiberg and E.~Witten,
  JHEP {\bf 9909} (1999) 032
  [arXiv:hep-th/9908142].




\bibitem{Kim:2002zg}
  N.~Kim and J.-H.~Park,
  Phys.\ Rev.\ D {\bf 66} (2002) 106007
  [arXiv:hep-th/0207061].

\bibitem{Dasgupta:2002ru}
  K.~Dasgupta, M.~M.~Sheikh-Jabbari and M.~Van Raamsdonk,
  JHEP {\bf 0209} (2002) 021
  [arXiv:hep-th/0207050].



\bibitem{Park:2002cb}
  J.-H.~Park,
  JHEP {\bf 0210} (2002) 032
  [arXiv:hep-th/0208161].



\bibitem{jhpBPS}
D.~Bak, K.~Lee and J.-H.~Park, Phys. Rev. D {\bf 66} (2002) 025021 [hep-th/0204221].



\bibitem{Kim:2002if}
  N.~Kim and J.~Plefka,
  Nucl.\ Phys.\ B {\bf 643} (2002) 31
  [arXiv:hep-th/0207034].





\bibitem{Kugo:1982bn}
  T.~Kugo and P.~K.~Townsend,
  Nucl.\ Phys.\ B {\bf 221} (1983) 357.



\bibitem{Park:1998nr}
J.-H.~Park,
Nucl.\ Phys.\ B {\bf 539} (1999) 599 [hep-th/9807186].




\bibitem{Kac:1977em}
  V.~G.~Kac,    ``Lie Superalgebras,''
  %
  Adv.\ Math.\  {\bf 26}, 8 (1977).

\bibitem{Kac:1977qb}
  V.~G.~Kac,   ``A Sketch Of Lie Superalgebra Theory,''
  %
  Commun.\ Math.\ Phys.\  {\bf 53}, 31 (1977).

\bibitem{Hyun:2002cm}
  S.~Hyun and H.~Shin,
  Phys.\ Lett.\ B {\bf 543} (2002) 115
  [arXiv:hep-th/0206090].




\bibitem{Lee:2004jx}
  S.~Lee and J.-H.~Park,
  JHEP {\bf 0406} (2004) 038
  [arXiv:hep-th/0404051].

\bibitem{Peeters:2003vz}
  K.~Peeters and M.~Zamaklar,
  Phys.\ Rev.\ D {\bf 69} (2004) 066009
  [arXiv:hep-th/0311110].



\bibitem{deWit:1998tk}
  B.~de Wit, K.~Peeters and J.~Plefka,
  %
  Nucl.\ Phys.\ B {\bf 532}, 99 (1998)
  [arXiv:hep-th/9803209].




\bibitem{Hyun:2002fk}
  S.~Hyun and J.-H.~Park,
  JHEP {\bf 0211} (2002) 001
  [arXiv:hep-th/0209219].


\bibitem{Hyun:2003se}
  S.~Hyun, J.-H.~Park and S.~H.~Yi,
  JHEP {\bf 0303} (2003) 004
  [arXiv:hep-th/0301090].



\end{thebibliography}
\end{document}